\documentclass[aps,superscriptaddress,floatfix,twocolumn,tightenlines,longbibliography]{revtex4-1}
\pdfoutput=1 

\usepackage{graphicx}
\usepackage{dcolumn}
\usepackage{bm}

\usepackage{bbm}
\usepackage{verbatim}
\usepackage{amsfonts}
\usepackage{amssymb}
\usepackage{amsmath}
\usepackage{amsthm}
\usepackage{graphicx}
\usepackage[bb=boondox]{mathalfa}
\usepackage{tensor}
\usepackage{times}
\usepackage[dvipsnames]{xcolor}
\usepackage{xspace}
\usepackage[normalem]{ulem}
\usepackage{subcaption}
\usepackage[unicode=true,pdfusetitle,
 bookmarks=true,bookmarksnumbered=false,bookmarksopen=false,
 breaklinks=false,pdfborder={0 0 1},backref=false,colorlinks=true,citecolor=blue,
 urlcolor=blue,linkcolor=blue]{hyperref}
\usepackage{physics}
\usepackage{tikz}
\usetikzlibrary{decorations.pathreplacing,calc,tikzmark}

\newcommand{\tkgroup}[1]{\hat{T}_k^{#1}}
\newcommand{\tkgeneral}{\tkgroup{G}}
\newcommand{\tkuhaar}{\tkgroup{U(\mathcal{H})}}
\newcommand{\tkunhaar}{\tkgroup{U(\mathcal{H}\vert\hat{N})}}
\newcommand{\tkughaar}{\tkgroup{U_b(\mathcal{H}\vert\hat{N})}}
\newcommand{\tkrandom}[1]{\hat{T}_k^{#1}}
\newcommand{\targrandom}[2]{\hat{T}_{#2}^{#1}}

\renewcommand{\vec}[1]{\boldsymbol{#1}}

\newcommand{\edgeket}[2]{{\ket{#1 ; #2}}}
\newcommand{\edgebra}[2]{{\bra{#1 ; #2}}}
\newcommand{\edgebraket}[4]{{\braket{#1; #2}{#3; #4}}}

\newcommand{\localdim}[2]{{d^{(#1)}_{#2}}}
\newcommand{\globaldim}[0]{{\mathcal{D}}}
\newcommand{\globaldimn}[1]{{d_{#1}}}

\newcommand{\unitaryaverage}[2]{{\int d#2 #1 }}

\newcommand{\kket}[1]{{\vert {#1} )}}
\newcommand{\bbra}[1]{{( {#1} \vert}}
\newcommand{\bbrakket}[2]{{( {#1} \vert {#2} )}}

\makeatletter
\newcommand{\cbigoplus}{\DOTSB\cbigoplus@\slimits@}
\newcommand{\cbigoplus@}{\mathop{\widehat{\bigoplus}}}
\makeatother

\begin{document}

\title{Unitary k-designs from random number-conserving quantum circuits}

\author{Sumner N. Hearth}
\author{Michael O. Flynn}
\author{Anushya Chandran}
\author{Chris R. Laumann}
\affiliation{%
Department of Physics, Boston University, 590 Commonwealth Avenue, Boston, Massachusetts 02215, USA
}
\date{\today}

\begin{abstract}
Local random circuits scramble efficiently and accordingly have a range of applications in quantum information and quantum dynamics.
With a global $U(1)$ charge however, the scrambling ability is reduced; for example, such random circuits do not generate the entire group of number-conserving unitaries.
We establish two results using the statistical mechanics of $k$-fold replicated circuits.
First, we show that finite moments cannot distinguish the ensemble that local random circuits generate from the Haar ensemble on the entire group of number-conserving unitaries.
Specifically, the circuits form a $k_c$-design with $k_c = O(L^d)$ for a system in $d$ spatial dimensions with linear dimension $L$.
Second, for $k < k_c$, we derive bounds on the depth $\tau$ required for the circuit to converge to an approximate $k$-design.
The depth is lower bounded by diffusion $k L^2 \ln(L) \lesssim \tau$.
In contrast, without number conservation $\tau \sim \text{poly}(k) L$. 
The convergence of the circuit ensemble is controlled by the low-energy properties of a frustration-free quantum statistical model which spontaneously breaks $k$ $U(1)$ symmetries.
We conjecture that the associated Goldstone modes set the spectral gap for arbitrary spatial and qudit dimensions, leading to an upper bound $\tau \lesssim k L^{d+2}$.

\end{abstract}

\maketitle

\section{Introduction}\label{sec:intro}

\begin{figure*}[t]
    \centering
    \begin{subfigure}{0.405\textwidth}
        \caption{}
        \label{fig:random_single_circuit}
        \includegraphics[width=\linewidth]{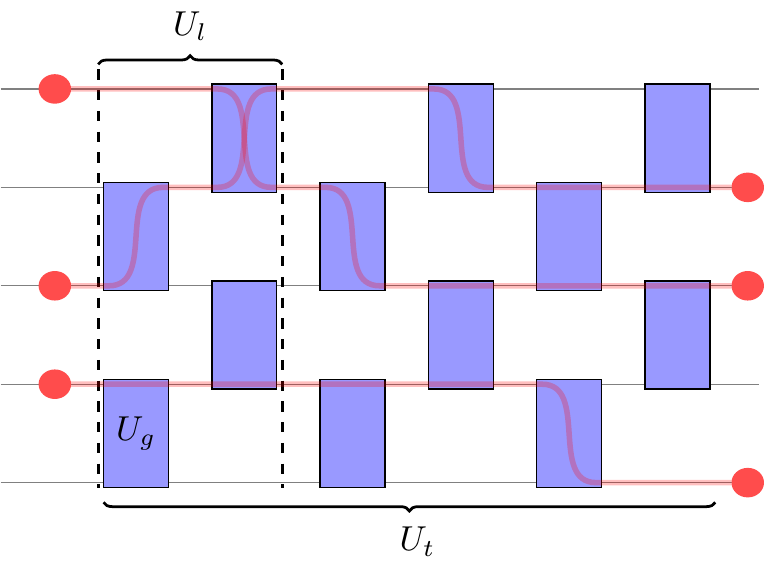}
    \end{subfigure}
    \hfill
    \begin{subfigure}{0.5\textwidth}
        \caption{}
        \label{fig:random_circuit}
        \includegraphics[width=\linewidth]{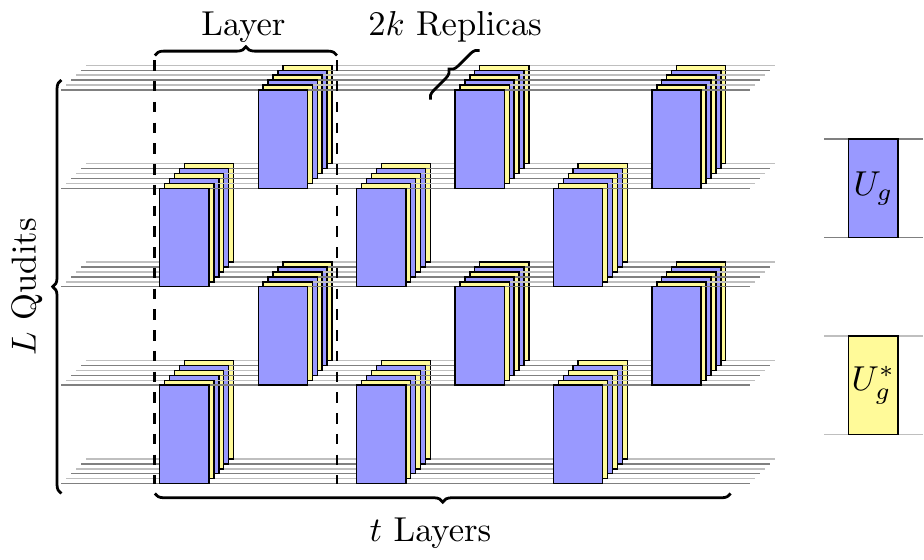}
    \end{subfigure}
    \caption{
    (a) A number-conserving brick-layer circuit in $d=1$ spatial dimensions with $L=5$ qudits, $t=3$ layers and $b=2$ body gates.
    The conserved number (red circles) spreads under the action of the circuit. 
    Each gate $U_g$ is independently and uniformly sampled from the set of local number-conserving unitaries.
    A layer unitary $U_l$ is the product of $U_g$ for each gate $g$ in a layer, and $U_t$ is the product of $t$ layers.
    (b) Replica model for the $k=3$ moment operator with $k=3$ replicas and $*$-replicas. 
    Each replica and $*$-replica carries a separately conserved number current.
    }
\end{figure*}

Consider a system of $L^d$ qudits arranged on a $d$-dimensional  lattice with a conserved $U(1)$ charge $\hat{N}$.
Let the qudits evolve according to a local unitary number-conserving circuit (see Fig.~\ref{fig:random_single_circuit}) .
When the gate unitaries are drawn uniformly at random, such circuits have a range of applications in quantum information and scrambling theory~\cite{knill2008,Harrow_2009, Hayden:2007aa,Brown:2015aa,fisher2022,McCulloch_DeNardis_Gopalakrishnan_Vasseur_2023,agrawal2022,Scott_2008,brown2013scrambling}.
For example, they provide minimal models for local unitary dynamics with conservation laws, permitting the study of operator hydrodynamics and scrambling~\cite{Rakovszky:2018aa,khemani2018a,gharibyan2018a,chan2018a,znidaric2020,rakovszky2021,fisher2022}.
They also enable randomized benchmarking protocols for measuring the purity of systems with number conservation such as ultracold atoms~\cite{vanenk2012,elben2018, brydges2019}, and produce output distributions from which it is difficult to sample~\cite{Boixo:2018aa,Bouland_2018,aaronson2020classical}. 

As the depth $t$ of the random circuit grows, the circuit unitary $U_t$ executes a random walk on the group of number-conserving unitaries $U(\mathcal{H}|\hat{N})$ acting on Hilbert space $\mathcal{H}$. 
The key question in the above applications boils down to how `scrambling' the circuit is -- for our purposes, how close the distribution of $U_t$ is to a steady-state Haar distribution on $U(\mathcal{H}|\hat{N})$. 
This can be quantified in several ways as these are very high dimensional distributions.
Luckily, physical applications typically only depend on low order moments of the circuit unitary.
These are naturally organized by the $k$'th \emph{moment operator},
\begin{align}
  \tkrandom{U_t} &\equiv \int dU \, p(U_t = U)  \left( U \otimes U^* \right)^{\otimes k} 
  \label{eq:momentopdef}
\end{align}
which encodes all of the $(k,k)$'th order moments~\footnote{Unbalanced moments in the matrix elements of $U_t$ and $U_t^*$ are encoded by $\targrandom{U_t}{k,\bar{k}}$. These can be non-zero for sufficiently large $k$, or $\bar{k}$ with number conservation, see Sec.~\ref{sec:moments_ng_unit_group}.}
in the matrix elements of $U_t$ and $U_t^*$.
Here, $p(U_t=U)$ is the probability distribution of the random variable $U_t$; for the Haar measure on a unitary group $G$, we also write $\tkgeneral$ and omit $p$ in a small abuse of notation.

If $\tkrandom{U_t} = \tkgeneral$ then the ensemble of random unitaries $U_t$ is said to form an \emph{exact $k$-design} on $G$ -- the distribution of $U_t$ is statistically indistinguishable from the Haar measure up to $k$'th order moments~\cite{roberts2017,gross2007}.
Typically, finite depth local circuits do not form exact designs; rather, they form \emph{approximate $k$-designs} which converge in the large depth limit
\begin{align}
\label{eq:conv_T_haar}
	\tkrandom{U_t} &\xrightarrow{t\to\infty} \tkgeneral.
\end{align}
In the absence of number conservation, for $G = U(\mathcal{H})$, this limit is well-studied~\cite{Harrow_2009,Diniz_2011,brandao2016a,brandao2016b,Haferkamp2022,Nakata_2017,harrow2023,hunter-jones2019,Brandao2021,HaferkampNature,Low:2010aa,jian2022,Emerson_2005,Gullans22,Schuster_Haferkamp_Huang_2024}.

Number conservation affects the limit, Eq.~\eqref{eq:conv_T_haar}, in two ways. 
First, the group generated by number-conserving $b$-body circuits of arbitrary depth, $U_b(\mathcal{H}|\hat{N})$, is a smooth proper subgroup of $U(\mathcal{H}|\hat{N})$ with codimension $O(L^d)$~\cite{marvian_restrictions_2022}. 
Thus, the moment operators for these two groups must differ for sufficiently large $k$. 
In particular, for arbitrary $k$, Eq.~\eqref{eq:conv_T_haar} holds for the Haar measure on $U_b(\mathcal{H}|\hat{N})$, not $U(\mathcal{H}|\hat{N})$. 
Second, charge transport is diffusive in a local number-conserving circuit. 
The depth required for convergence should thus grow at least as quickly as $L^2$. 
In contrast, $b$-body circuits without conservation laws have been shown to converge to approximate designs on the full unitary group $U(\mathcal{H})$ in depths of order $L$ in $d=1$~\cite{Harrow_2009,Diniz_2011,brandao2016a,Nakata_2017,harrow2023,hunter-jones2019}. 
More recent work has demonstrated convergence in extremely low depths of $\text{poly} \log(L)$~\cite{Schuster_Haferkamp_Huang_2024}.

We present two main results in this article. 
The first, in Sec.~\ref{sec:moments_ng_unit_group}, is that all finite moments are identical for the Haar measures on $U_b(\mathcal{H}|\hat{N})$ and $U(\mathcal{H}|\hat{N})$ as $L\to \infty$. 
More precisely, we show that for $b$-body circuits with $b\ge2$,
\begin{align}
\label{eq:T_haar_Ug_U}
	\tkughaar &= \tkunhaar, \quad k<k_c
\end{align}
with $k_c \ge L^d$. 
Finite moments \emph{cannot} detect that $b$-body circuits diffuse on $U_b(\mathcal{H}|\hat{N})$ rather than $U(\mathcal{H}|\hat{N})$ in the thermodynamic limit.

\begin{figure*}
    \centering
    \begin{subfigure}{0.25\textwidth}
        \caption{}
        \label{fig:HFF-spec-noncon}
        \includegraphics[width=\linewidth]{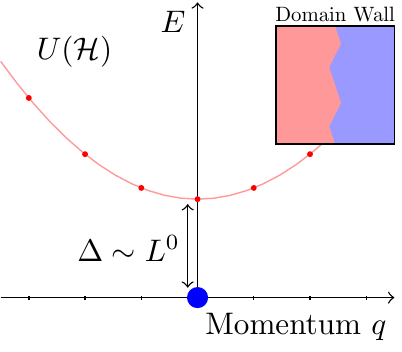}
    \end{subfigure}
    \qquad
    \begin{subfigure}{0.25\textwidth}
        \caption{}
        \label{fig:HFF-spec-con}
        \includegraphics[width=\linewidth]{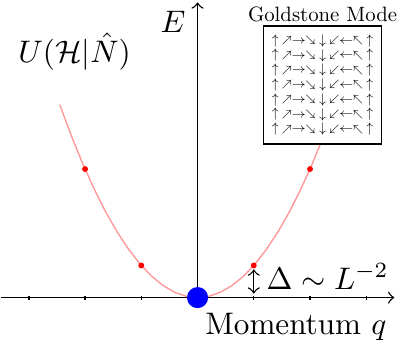}
    \end{subfigure}
    \qquad
    \begin{subfigure}{0.275\textwidth}
        \caption{}
        \label{fig:convergenceplot}
        \includegraphics[width=\linewidth]{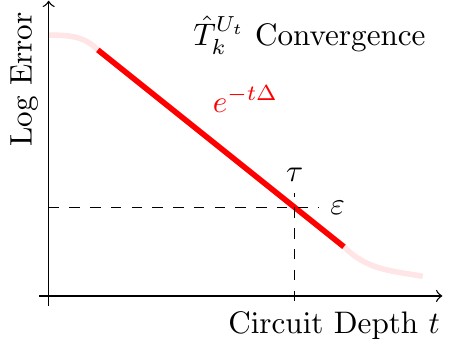}
    \end{subfigure}
    \caption{
    The low energy spectrum of the replica model (a) without and (b) with number conservation.
    Dashes on the x-axis mark the discrete set of momenta at finite size, and the red dots mark the corresponding excitation energies.
    Without number conservation, the ground state spontaneously breaks the $S_k \times S_k$ symmetry down to a diagonal subgroup $S_k$. 
    The domain wall excitations are gapped.
    With number conservation, the ground state additionally spontaneously breaks the $U(1)^k \times U(1)^k$ symmetry down to a diagonal $U(1)^k$ subgroup. 
    The resulting $k$ gapless Goldstone modes having dynamical exponent $z=2$, owing to the frustration-freeness of the model.
    This leads to a finite-size gap $\Delta \sim L^{-2}$.
    (c) The spectral gap $\Delta$ sets the timescale for convergence to an approximate $k$-design. 
    The convergence time $\tau$ is the depth required to reach an error, or diamond norm difference, of $\varepsilon$ (Sec.~\ref{sec:kdesignconvergencetime})
    }
    \label{fig:HFF-spec}
\end{figure*}

Second, in Sec.~\ref{sec:circuits}, we show that $\tkrandom{U_t}$ converges to $\tkunhaar$ exponentially with a rate $\Delta$ limited by diffusion,
\begin{align}
\label{eq:t_conv}
    \tkrandom{U_t} &\approx \tkunhaar + \hat{C} e^{-\Delta t} + \cdots, \qquad \Delta \lesssim \frac{1}{L^2}
\end{align}
This follows from interpreting $\tkrandom{U_t}$ as a replicated quantum statistical mechanical model (see Fig.~\ref{fig:random_circuit}) whose low energy properties govern the late time convergence~\cite{Brown_Viola_2010,Chan_De_Luca_Chalker_2018,nahum_operator_2018,zhou2019,Barratt:2022aa,Agrawal_Zabalo_Chen_Wilson_Potter_Pixley_Gopalakrishnan_Vasseur_2022}.
The frustration-free ground state of the replicated model spontaneously breaks $k$ $U(1)$ symmetries. 
The rate $\Delta$ in Eq.~\eqref{eq:t_conv} is thus bounded by the gap to the lowest wavenumber Goldstone mode (Fig.~\ref{fig:HFF-spec-con}).

We use this bound on $\Delta$ to rigorously bound $\tau$, the time at which the circuit converges to an $\varepsilon$-approximate $k$-design on  $U(\mathcal{H}|\hat{N})$ (See Fig.~\ref{fig:convergenceplot}), from below.
With $M\propto L^d$ number sectors, each converging on a timescale of $1/\Delta$, we find
\begin{align}
\label{eq:tau_lowerbound}
    \tau \gtrsim k d L^2 \ln(L) - L^2\ln(\varepsilon) + O(L^{2-kd}),
\end{align}
where the $\sim$ hides geometry-dependent constants.
If we further conjecture that the Goldstone modes are the lowest lying excitations and determine $\Delta$, we find that random number-conserving circuits form $\varepsilon$-approximate $k$-designs at a depth
\begin{align}
    \tau \lesssim k L^{d+2} \ln(Q) - L^2\ln(\varepsilon).
\end{align}
Here, $Q$ is the total Hilbert space dimension of each qudit.

The paper proceeds as follows. 
Sec.~\ref{sec:the_main_event} presents the high-level spontaneous symmetry breaking analysis that leads to Eq.~\eqref{eq:t_conv}.  
Sec.~\ref{sec:notation} collects definitions and symbols used throughout the paper. 
Sec.~\ref{sec:ReviewHaar} reviews general properties of the Haar measure and the moment operators for number-conserving unitary groups.
Sec.~\ref{sec:moments_ng_unit_group} presents the first of our main results: all finite moments of $U_{b}(\mathcal{H}|\hat{N})$ and $U(\mathcal{H}|\hat{N})$ are identical in the thermodynamic limit. 
Sec.~\ref{sec:circuits} analyzes the low energy properties of the replicated circuit model and an associated Hamiltonian model, both of which are frustration-free. 
There we prove our second main result: $\tkrandom{U_{t}}$ converges exponentially to $\tkunhaar$ with a rate $\Delta\sim 1/L^{2}$. 
Sec.~\ref{sec:kdesignconvergencetime} translates the convergence rate $\Delta$ into a bound on the depth, $\tau$, beyond which the circuit is an approximate unitary $k$-design. 
We conclude with a discussion in Sec.~\ref{sec:Discussion}.

\section{Overview of Replica Model}
\label{sec:the_main_event}

In what follows, we present the arguments that lead to the second main result, Eq.~\eqref{eq:t_conv}, in more detail. 
The moment operator for $U_t$ is a product of moment operators on layers, each of which is product of moment operators on gates,
\begin{align}
\label{eq:tprod}
	\tkrandom{U_t} 
	&= \prod_{\mathrm{Layers~} l} \underbrace{\left(\prod_{\text{Gates }g\in l} \tkrandom{U_g}
 \right)}_{\tkrandom{U_l}} = \left(\tkrandom{U_l}\right)^t
\end{align}
Clearly, $\tkrandom{U_l}$ plays the role of a transfer matrix in the treatment of the replicated system as a quantum statistical mechanical model in $d+1$ dimensions~\cite{Brown_Viola_2010,Chan_De_Luca_Chalker_2018,nahum_operator_2018,zhou2019,Barratt:2022aa,Agrawal_Zabalo_Chen_Wilson_Potter_Pixley_Gopalakrishnan_Vasseur_2022}.
The convergence of $\tkrandom{U_t}$ is thus governed by the maximum eigenvalue of $\tkrandom{U_l}$ and the spectral gap below it. 

To make progress, we make several observations.
First, each gate operator, $\tkrandom{U_g}$, is a Hermitian projector, with eigenvalues $+1$ and $0$. 
This elementary result holds for the moments of any Haar distributed unitary (see Sec.~\ref{sec:ReviewHaar} for a review of general properties of the moment operator for Haar ensembles). 
As a product of projectors, the maximum possible eigenvalue of the transfer matrix $\tkrandom{U_l}$ is $+1$, carried by steady states which are simultaneous $+1$ eigenstates of \emph{all} of the gate projectors $\tkrandom{U_g}$. 
So long as this joint steady state space is non-empty, $\tkrandom{U_t}$ converges to a finite projector, as expected in Eq.~\eqref{eq:conv_T_haar}. 

While it is possible to further analyze $\tkrandom{U_l}$ directly (see Sec.~\ref{ssec:circuit_gap}), it is instructive to consider the closely related layer Hamiltonian of the replicated system,
\begin{align}
    \hat{H}_k &= \sum_{\text{Gates~}g \in l} (1 - \tkrandom{U_g})
    \label{Eq:HamFF}
\end{align}
This Hamiltonian is \emph{frustration-free}: its zero energy ground state space is the joint zero energy ground state space of each term -- that is, the ground state space of $\hat{H}_k$ is the steady state space of $\tkrandom{U_l}$.
Technically, there are two advantages to shifting our attention to the Hamiltonian $\hat{H}_k$ rather than $\tkrandom{U_l}$. 
First, while $\hat{H}_k$ has all of the spatial structure of the gates in a circuit layer, it does not keep track of their ordering between sublayers as in Fig.~\ref{fig:random_circuit}, which simplifies its analysis.
We do not expect this to qualitatively matter to physical properties of the low temperature limit, where the layers are iterated many times; indeed, we find it does not.
Second, it leads us to interpret the steady states of the replicated system in terms of spontaneous symmetry breaking in the ground states of the frustration-free Hamiltonian $\hat{H}_k$.

Let us then turn to the symmetries of the replicated system.
As in all replica treatments, the replicated system has $S_k \times S_k$ replica (and $*$-replica) permutation symmetry.
In the case of unitary circuits without symmetries, it is well known that the replicated system can be mapped to a ferromagnet with sites labeled by elements of $S_k$ ~\cite{Hayden_Nezami_Qi_Thomas_Walter_Yang_2016,zhou2019}.
In this sense the model spontaneously breaks the $S_k \times S_k$ symmetry down to a diagonal subgroup $S_k$. 
More precisely, the ground state space of $\hat{H}_k$ is spanned by states labeled by permutations $\sigma\in S_{k}$ which bind the state on replica $\alpha$ to that on $*$-replica $\sigma(\alpha)$,
\begin{align}
    \ket{\sigma} &= \sum_{\vec{i},\bar{\vec{i}}} \underbrace{\left(\prod_{\alpha=1}^k \delta[\bar{i}_{\alpha} = i_{\sigma(\alpha)}]\right)}_{\delta[\bar{\vec{i}} = \sigma(\vec{i})]} \ket{\vec{i},\bar{\vec{i}}} = \sum_{\vec{i}} \ket{\vec{i},\sigma(\vec{i})} \label{eq:sigmastates_intro} 
\end{align}
where $i_\alpha$ ($\bar{i}_\alpha$) runs over basis states for the $\alpha$ replica ($*$-replica).
In the simplest case, when the dimension of the Hilbert space that each gate acts on is larger than $k$, these states are linearly independent and we obtain a ground state space with dimension $k!$.
As a discrete symmetry breaking phase in a lattice system, the Hamiltonian $\hat{H}_k$ has a spectral gap $\Delta \sim O(L^0)$ for large $L$; accordingly, $\tkrandom{U_t}$ converges exponentially with an order one timescale. The lowest lying states above the gap are well-described by single domain wall excitations, see Fig.~\ref{fig:HFF-spec-noncon}.

Returning to the case of a number-conserving circuit, each replica (and $*$-replica) additionally carries its own conserved charge associated to the replicated $U(1)$ symmetry -- the system has a global $U(1)^k \times U(1)^k$ symmetry.
This breaks spontaneously down to a diagonal subgroup $U(1)^k$, which is intertwined with the choice of replica permutation symmetry breaking. 
Explicitly, the ground states of $\hat{H}_k$ are spanned by the symmetry-breaking vacua,
\begin{align}
    \edgeket{\vec{\Phi}}{\sigma} &= e^{i\sum_\alpha \Phi_\alpha \hat{N}_\alpha } \ket{\sigma} \label{eq:phasestates}
\end{align}
where $\vec{\Phi}$ is a collection of $k$ phases, one to each replica.
Although the parameters $\vec{\Phi}$ vary continuously, they span a ground state space of dimension $k! M^k$ where $M$ is the number of global charge sectors in the system (again in the simplest case for small enough $k$); indeed, a basis for these states may be obtained by projecting the parent $\edgeket{\vec{\Phi}}{\sigma}$ states into the global charge sectors (Sec.~\ref{ssec:moments_ninv_unit_group}).

As $\hat{H}_k$ spontaneously breaks a continuous symmetry, it harbors $k$ Goldstone modes associated with long-wavelength fluctuations of the phases $\vec{\Phi}$, see Fig.~\ref{fig:HFF-spec-con}.
Such excitations can be created with momentum $q$ above any symmetry breaking vacuum using the modulated number operator,
\begin{align}
    \hat{n}^{\alpha}_{q} &= \sum_{x} e^{-i q \cdot x} \hat{n}^{\alpha}_{x} 
\end{align}
where $\hat{n}^{\alpha}_{x}$ is the local number operator at qudit $x$ in replica $\alpha$.
The excitation energy of these modes provides an upper bound on the spectral gap $\Delta $ of $\hat{H}_k$.
Indeed, in Sec.~\ref{ssec:VariationalHam} we find $\Delta \lesssim \frac{1}{L^2}$, as expected for Goldstone modes in frustration-free systems~\cite{Gosset_Mozgunov_2016,Masaoka_Soejima_Watanabe_2024,Masaoka_Soejima_Watanabe_2024_2,Lemm_Lucia_2024}, independent of $k$, spatial dimension $d$, and local Hilbert space structure.

The frustration-free Hamiltonian $\hat{H}_k$ has been previously studied in two special cases in $d=1$~\cite{gharibyan2018a,Barratt:2022aa}. 
In Ref.~\cite{gharibyan2018a}, the qudit on each site is composed of a charged qubit and a neutral qudit with an infinitely large Hilbert space dimension. 
Ref.~\cite{Barratt:2022aa} considers the case of qubits on each site. In both cases, $\hat{H}_k$ is a nearest neighbor Heisenberg model in each replica. 
The nearest neighbor Heisenberg model is special because it has an enlarged $SU(2)$ symmetry and is integrable in one dimension. 
Indeed, Ref.~\cite{gharibyan2018a} obtained the spectral gap from the integrability literature. 
Our work shows that $\hat{H}_k$ can be derived for \emph{any} local qudit structure in any dimension, and in general, need not be integrable or have an enlarged symmetry. 
It is however frustration-free and spontaneously breaks the replicated $U(1)$-symmetry, which guarantees $\Delta \lesssim 1/L^2$. 

Although we have sketched the arguments here for the associated Hamiltonian system, Eq.~\eqref{Eq:HamFF}, the physics is the same in the discrete time system defined by Eq.~\eqref{eq:tprod}.
Sec.~\ref{ssec:circuit_gap} deals with the mathematical tweaks required to analyze the discrete time case.

\section{Definitions and Notation}
\label{sec:notation}

We consider a system of $V = L^d$ qudits in $d$ spatial dimensions with linear dimension $L$. 
Under the $U(1)$ symmetry, each qudit's local Hilbert space $\mathcal{H}^{(1)}$ can be decomposed into $M^{(1)}$ number sectors,
\begin{align}
     \mathcal{H}^{(1)} &= \bigoplus_{n=0}^{M^{(1)}-1}\mathcal{H}^{(1)}_n.
\end{align} 
We denote the local Hilbert space dimension of number sector $n$ as $\localdim{1}{n} = |\mathcal{H}^{(1)}_n|$.
For example, in a lattice model of single-orbital spin-$1/2$ fermions, the local Hilbert space may have $0,1$ or $2$ particles and we write $d^{(1)} = (1,2,1)$. 
In a spin-$1/2$ system with $S^z$ conservation, $d^{(1)} = (1,1)$.

The local dimensions $d^{(1)}$ fully specify the decomposition of the multi-qudit Hilbert space.
For example, a pair of qudits has joint Hilbert space $\mathcal{H}^{(2)}$, which can be similarly decomposed into number sectors
\begin{align}
    \mathcal{H}^{(2)} = \mathcal{H}^{(1)}\otimes\mathcal{H}^{(1)} &= \sum_{n=0}^{M^{(2)} - 1} \mathcal{H}^{(2)}_n
\end{align}
\begin{align}
    \localdim{2}{n} \equiv |\mathcal{H}^{(2)}_n| = \sum_{n_1 + n_2 = n} \localdim{1}{n_1} \localdim{1}{n_2} . \label{eq:localhilbertspacesector}
\end{align}
The Hilbert space of the $V$ qudits is denoted by $\mathcal{H} = \mathcal{H}^{(V)}$, with $\globaldim = \vert \mathcal{H} \vert$ (we omit sub/superscripts to denote properties of a complete set of qudits).
$\mathcal{H}$ and $\globaldim$ may in turn be broken into global number sectors as $\mathcal{H} = \sum_n \mathcal{H}_n$ and $\globaldim = \sum_n d_n$ with $d_n = \vert \mathcal{H}_n \vert$.

We take the standard basis $\ket{i}$ to be compatible with the decomposition of $\mathcal{H}$ into number sectors and qudits. 
That is, we write the global number operator
\begin{align}
    \hat{N} = \sum_x \hat{n}_x
\end{align}
where $x$ runs over spatial sites. 
Then, the site-resolved number operators are diagonal
\begin{align}
    \hat{n}_x\ket{i} = n_x(i) \ket{i}.
    \label{eq:standardbasis}
\end{align}

We consider circuits with brick-layer architectures as in Fig.~\ref{fig:random_circuit}. 
Each of the $t$ layers is obtained by stacking a finite number of sub-layers of gates acting on all even bonds along each spatial dimension, and odd bonds along each spatial dimension. 
The ordering of the sub-layers is not important. 
Thus, for example, in $d=1$, each layer is composed of two sub-layers (Fig.~\ref{fig:random_circuit}), while in $d=2$, there are four sub-layers and the resulting interaction graph is that of a square lattice. 
The total number of layers is denoted by $t$. 
Although the geometry of each layer is repeated, the particular gates $U_g$ are not -- they are each independently sampled from the uniform measure on the group of number-conserving 2-body gates.

Table~\ref{tab:symbol_glossary} provides a comprehensive symbol glossary.

\newcommand\Tstrut{\rule{0pt}{2.6ex}}         
\newcommand\Bstrut{\rule[-0.9ex]{0pt}{0pt}}   

\begin{table*}[t]
    \centering
    \begin{tabular}{l|c|c}
    Name & Symbol & Notes  \\ 
    \hline\hline\Tstrut
    Number of qudits & $V = L^d$ &  \\
    Number of global number sectors & $M$ & \\
    Spatial dimension & $d$ & \\
    Number of qudits acted on by a gate & $b$ & \\ 
    \hline \Tstrut
    Dimension of $n$th number sector for $m$ sites & $\localdim{m}{n}$ & Eq.~\eqref{eq:localhilbertspacesector} \\
    Dimension of $n$th global number sector & $\globaldimn{n}$ & $\globaldimn{n} = \localdim{V}{n}$ \\
    Single-site Hilbert space dimension & $Q$ & $Q = \sum_n \localdim{1}{n}$\\
    Global Hilbert space dimension & $\globaldim$ & $\globaldim = Q^V = \sum_n \globaldimn{n}$ \\ 
    \hline \Tstrut
    Replica index & $\alpha $ & $\alpha \in 1 \cdots k$\\
    Qudit index & $x$ & $x \in 1 \cdots V$ \\
    Number sector index & $n$ & $n \in 0 \cdots M-1$  \\
    \hline \Tstrut 
    Number-conserving unitary group & ~$U(\mathcal{H}|\hat{N})$~ & Eq.~\eqref{eq:nconservingugroup}\\
    Unitaries realizable from arbitrary depth $b$-body circuits & ~$U_b(\mathcal{H}|\hat{N})$~ & ~ Sec.~\ref{sec:moments_ng_unit_group}\\
    \hline\Tstrut
    Moment operator for random unitary $U$ & $\tkrandom{U}$ & Eq.~\eqref{eq:momentopdef}  \\
    Moment operator for Haar measure on group $G$ & $\tkrandom{G}$ & Eq.~\eqref{eq:momentoperatorhaargroup} \\
    \hline  \Tstrut
    Moment operator for layer/transfer matrix & $\tkrandom{U_l}$ & Eq.~\eqref{eq:tprod}\\
    Hamiltonian of associated replica model & $\hat{H}_k$ & Eq.~\eqref{Eq:HamFF2} \\
    Gap of replica model& $\Delta$ & Eq.~\eqref{eq:t_conv}\\
    Time to converge to approximate design & $\tau$ & Sec.~\ref{sec:kdesignconvergencetime}\\
    \hline \Tstrut
    Standard basis of Hilbert space $\mathcal{H}$ & $\ket{i}$ & Sec.~\ref{sec:notation} \\
    Standard basis of replica space $\left(\mathcal{H}\otimes\mathcal{H}^*\right)^{\otimes k}$ & $\ket{\vec{i}, \vec{\bar{i}}}$ & Sec.~\ref{sec:notation} \\
    Permutation state & $\ket{\sigma}$ & Eq.~\eqref{eq:sigmastates} \\
    Number-permutation state & $\edgeket{\vec{n}}{\sigma}$ & Eq.~\eqref{eq:numberstates} \\
    Phase-permutation state & $\edgeket{\vec{\Phi}}{\sigma}$ & Eq.~\eqref{eq:phasestates}
    \end{tabular}
    \caption{%
    Symbol glossary. Vectors (eg. $\vec{i}$) have components across $k$ replicas ($i_\alpha$). Bar indicates $*$-replicas.
    }
    \label{tab:symbol_glossary}
\end{table*}

\section{Moments of Global Unitary Groups}
\label{sec:ReviewHaar}

We briefly review key properties of the Haar ensemble on compact unitary (matrix) groups $G$ acting on a Hilbert space $\mathcal{H}$. 
We begin with features shared by all compact unitary groups and then turn to the cases of the general unitary group $U(\mathcal{H})$ and the number-conserving unitary group $U(\mathcal{H}|\hat{N})$.
The results on $U(\mathcal{H}|\hat{N})$ apply directly to the individual gates in the local circuit model.

\subsection{General Properties}
\label{ssec:generalproperties}

Suppose $G$ is a compact group of unitaries acting on the Hilbert space $\mathcal{H}$ with dimension $\globaldim$.
This could be the general unitary group $U(\mathcal{H})$ or some compact subgroup of it, such as $U(\mathcal{H}|\hat{N})$.
The Haar measure on $G$ is the unique normalized measure which is invariant under both left- and right- shifts, as well as inversion~\cite{HaarJoy}. 
That is, for arbitrary functions $f$ on $G$ and elements $U, V \in G$, 
\begin{align}
\label{eq:haarinvprops}\hspace{-4pt}
    \unitaryaverage{f(U)}{U} = \hspace{-1pt} \unitaryaverage{f(VU)}{U} \hspace{-1pt} = \unitaryaverage{f(UV)}{U} = \hspace{-1pt} \unitaryaverage{f(U^{-1})}{U}
\end{align}
These invariance properties completely characterize the measure.

\paragraph{The Moment Operator---} 
Moments of $U$ are encoded in the moment operator 
\begin{align}
    \targrandom{G}{k,\bar{k}} &\equiv \unitaryaverage{\left(U^{\otimes k} \otimes (U^*)^{\otimes \bar{k}}\right)}{U} 
    \label{eq:momentoperatorhaargroup}
\end{align}
which acts on the $(k+\bar{k})$-fold replicated Hilbert space $\mathcal{H}^{\otimes k} \otimes \left(\mathcal{H}^*\right)^{\otimes \bar{k}}$. 

When the number of starred and unstarred replicas are unequal, $k \neq \bar{k}$, the moment operator is called unbalanced. In many cases, the unbalanced moment operators are identically zero. 
Thus, we largely focus on the balanced moment operators, $\tkgeneral \equiv \targrandom{G}{k,k}$ and only discuss unbalanced moments when we need to show that they vanish (see Sec.~\ref{ssec:moments_gen_unitary_group}).

The matrix elements of the moment operator encode the $\globaldim^{4k}$ moments of order $(k,k)$ of the matrix elements of $U$ and $U^*$.
For arbitrary compact unitary groups $G$, $\tkgeneral$ is a \emph{Hermitian projector} -- a fact that follows directly from the unitarity of $G$ and the invariance properties in Eq.~\eqref{eq:haarinvprops}. 
Explicitly, shift-invariance implies
\begin{equation}
    \begin{aligned}
        \left(\tkgeneral\right)^2 &=  \int dU dV \left(UV \otimes U^* V^*\right)^{\otimes k}\\
        &= \left(\int dV \right) \tkgeneral\\
        &= \tkgeneral .
    \end{aligned} \label{eq:t_projector}
\end{equation}
while inversion-invariance gives
\begin{equation}
    \begin{aligned}
        \left(\tkgeneral\right)^\dagger &= \left(\int dU (U \otimes U^*)^{\otimes k}\right)^\dagger\\
        &= \int dU (U^{-1} \otimes U^{-1*})^{\otimes k}\\
        &= \tkgeneral .
    \end{aligned} \label{eq:t_hermitian}
\end{equation}
Eq.~\eqref{eq:t_projector} and \eqref{eq:t_hermitian} together imply $\tkgeneral$ is a Hermitian projector, with eigenvalues of $+1$ and $0$.

As a projector, $\tkgeneral$ is completely characterized by its $+1$ eigenspace.
Below, we construct these spaces explicitly for the Haar measure on several unitary groups.
First, we review a few useful general relationships. 

\paragraph{Restricted Measures and Subgroups---} \label{par:restricted_subgroups}
Suppose unitary $V$ is sampled according to an arbitrary non-Haar measure on $G$. 
The moment operator for $V$, $\tkrandom{V}$, need not be a projector. 
Nonetheless, shift-invariance of the Haar measure implies that $\tkrandom{V} \tkgeneral = \tkgeneral \tkrandom{V} = \tkgeneral$. 
Less abstractly, the $+1$ eigenstates of the Haar moment projector are also $+1$ eigenstates of $\tkrandom{V}$.

In particular, if $V$ is sampled from the Haar measure on a compact subgroup $H \subset G$, $\tkrandom{V}$ projects onto a superspace of $\tkgeneral$. 
This is useful for constructing the $+1$ eigenspace of restricted unitary groups like $U(\mathcal{H}|\hat{N})$, as we will see in Sec.~\ref{ssec:moments_ninv_unit_group}. 

\paragraph{Replica Symmetry---}
From its definition, the moment operator $\tkgeneral$ is clearly symmetric under the $S_k \cross S_k$ separate permutations of replicas and $*$-replicas.
Indeed, replica permutation symmetry holds for any ensemble on $G$.
To make this explicit and set some notation, let us introduce an orthonormal basis for the $k$ replicas labeled by $\vec{i} = (i_1 \cdots i_k)$, and similarly for the $k$ $*$-replicas $\bar{\vec{i}} = (\bar{i}_1 \cdots \bar{i}_k)$, where each $i_\alpha, \bar{i}_\alpha =1, \dots \globaldim$.
$\tkgeneral$ acts on states $\ket{\vec{i},\bar{\vec{i}}} \in \left(\mathcal{H}\otimes\mathcal{H}^*\right)^{\otimes k}$.
Consider permutations $\sigma, \tau \in S_k$ and associated operator $\hat{R}_{(\sigma,\tau)} = \sum_{\vec{i},\bar{\vec{i}}} \ket{\sigma(\vec{i}),\tau(\bar{\vec{i}})}\bra{\vec{i},\bar{\vec{i}}}$. 
The action on $\tkgeneral$ can be understood,
\begin{align}
    \bra{\vec{j},\bar{\vec{j}}} \hat{R}_{(\sigma,\tau)}^{-1} \tkgeneral &\hat{R}_{(\sigma,\tau)} \ket{\vec{i},\bar{\vec{i}}} \nonumber  \\ &= \int dU \prod_\alpha U_{i_{\sigma(\alpha)}j_{\sigma(\alpha)}} \prod_\beta U^*_{\bar{i}_{\tau(\beta)}\bar{j}_{\tau(\beta)}} \nonumber \\
    &= \bra{\vec{j},\bar{\vec{j}}} \tkgeneral \ket{\vec{i},\bar{\vec{i}}} . \label{eq:permutationsymmetryproof}
\end{align}
Rearranging the matrix elements in the integrand has no effect on the ensemble average, and the moment operator is invariant under replica and $*$-replica permutations.

\paragraph{Associated Hamiltonian---}
Replica symmetry implies that the $+1$ eigenspace must be invariant under $S_k \times S_k$ permutations, though the action inside that space may be non-trivial. 
Indeed, a useful perspective is provided by considering the replica-symmetric Hamiltonian  whose ground state space is the $+1$ eigenspace of $\tkgeneral$,
\begin{align}
    \hat{H}_k = 1 - \tkgeneral
\end{align}
If the subgroup $\mathcal{S} \subseteq S_k \times S_k$ leaves a particular ground state of $\hat{H}_k$ invariant, then we can say that the $S_k \times S_k$ symmetry is spontaneously broken down to $\mathcal{S}$.

\paragraph{Spatial permutation symmetry---}
Suppose the Hilbert space $\mathcal{H}$ is a tensor product of single qudit Hilbert spaces, so that the moment operator has $2kV$ incoming and outgoing qudit `legs', $V$ per replica and $*$-replica.
Let the operator $\hat{L}_\pi = \hat{l}_\pi^{\otimes 2k}$ apply the same permutation $\pi \in S_V$ to the spatial indices in each replica. 
Left- and right- shift invariance of the Haar measure implies that the moment operator has left- and right- leg permutation symmetry,
\begin{align}
\tkgeneral \hat{L}_\pi = \hat{L}_\pi \tkgeneral = \tkgeneral. \label{eq:tlegsym}
\end{align}

\subsection{General Unitary Group $U(\mathcal{H})$}
\label{ssec:moments_gen_unitary_group}

Let us now focus on the moments $\tkuhaar$ for the general unitary group $U(\mathcal{H})$ without number conservation. 
As a consequence of Schur-Weyl duality~\cite{roberts2017} (see App.~\ref{app:Schurweyl}), $\tkuhaar$ projects onto the space spanned by the $k!$ permutation states,
\begin{align}
    \ket{\sigma} &= \sum_{\vec{i},\bar{\vec{i}}} \underbrace{\left(\prod_{\alpha=1}^k \delta[\bar{i}_{\alpha} = i_{\sigma(\alpha)}]\right)}_{\delta[\bar{\vec{i}} = \sigma(\vec{i})]} \ket{\vec{i},\bar{\vec{i}}} = \sum_{\vec{i}} \ket{\vec{i},\sigma(\vec{i})} \label{eq:sigmastates} 
\end{align}
which bind replicas to $*$-replicas. 
Accordingly, $\tkuhaar$ has an operator expansion of the form
\begin{align}
    \tkuhaar &= \sum_{\sigma,\tau \in S_k}  V^{\sigma\tau} \ketbra{\sigma}{\tau} .
    \label{eq:tkexpansion}
\end{align}
Pictorially (for $k=3$),
\begin{align}
    \begin{tikzpicture}[scale=0.4,baseline={([yshift=-.5ex]current bounding box.center)}]
        \foreach \j in {5,...,0}{
            \draw (-0.25,\j+0.5) -- (5.25,\j+0.5);
        }
        \draw[fill=blue!10] (1,-0.25) rectangle (4,6.25);
        \node at (2.5,3) {$\tkuhaar$};
    \end{tikzpicture} &= \sum_{{\color{red}\sigma},{\color{blue}\tau} \in S_k} V^{{\color{red}\sigma}{\color{blue}\tau}} \left(
    \begin{tikzpicture}[scale=0.4,baseline={([yshift=-.5ex]current bounding box.center)}]
        \foreach \j in {5,...,0}{
            \draw (-0.25,\j+0.5) -- (1,\j+0.5);
            \draw (4,\j+0.5) -- (5.25,\j+0.5);
        }
        \draw[fill=blue,opacity=0.05] (1,-0.25) rectangle (4,6.25);
        \foreach \j in {2,...,0}{
            \draw[red] (1,2*\j+0.5) -- (1.75,2*\j+0.5) -- (1.75,2*\j+1+0.5) --(1,2*\j+1+0.5);
        }
        \draw[blue] (4,2*2+0.5) -- (3.25,2*2+0.5) -- (3.25,2*2+1+0.5) --(4,2*2+1+0.5);
        \draw[blue] (4,2*0+0.5) -- (3.25,2*0+0.5) -- (3.25,2*1+1+0.5) --(4,2*1+1+0.5);
        \draw[blue] (4,2*0+1+0.5) -- (3.75,2*0+1+0.5) -- (3.75,2*1+0.5) --(4,2*1+0.5);
    \end{tikzpicture}\right) .
\end{align}

The coefficients $V^{\sigma\tau}$, better known as the Weingarten functions \cite{Weingarten1978}, are determined by the overlaps of the states $\ket{\sigma}$ onto which $\tkuhaar$ projects.  
These are neither orthogonal nor normalized -- indeed, they are not linearly independent for $k>\globaldim$. 
The formulae for $V^{\sigma\tau}$ are presented in Appendix~\ref{App:Vsigmatau} for any $k$ for completeness, but will not be needed in the main text.
Furthermore, the states $\ket{\sigma}$ are linearly independent only if $k \le \globaldim$. 
If $k$ is too large we must instead draw from a linearly independent set of states labeled by $\sigma$ in the set $S_k^{\le \globaldim} \subseteq S_k$. 
The details of this set are left to Appendix~\ref{App:Vsigmatau}.
Rather than clutter the already burdened notation, we leave the set from which permutations are drawn implicit.

The states $\ket{\sigma}$ transform non-trivially under the $S_k\cross S_k$ replica symmetry (c.f. Eq.~\eqref{eq:permutationsymmetryproof}):
\begin{align}
\hat{R}_{(\sigma,\tau)} \ket{\gamma} = \ket{\sigma^{-1} \tau \gamma }
\end{align}
That is, the space spanned by $\{\ket{\sigma}\}$ is invariant under replica symmetry $S_k \times S_k$ (as it had to be), but any given state $\ket{\sigma}$ is invariant only under a subgroup isomorphic to $S_k$. 
In the language of the ground state space of the Hamiltonian $\hat{H}_k = 1-\tkuhaar$, we  say that the $S_k \times S_k$ replica symmetry is spontaneously broken down to the diagonal subgroup $S_k$.

The unbalanced moment operators are identically zero, a simple consequence of shift invariance. Consider left-multiplying by the unitary $V = e^{i \theta}$. 
\begin{align}
V^{\otimes k} (V^*)^{\otimes \bar{k}} \targrandom{U(\mathcal{H})}{k, \bar{k}} &= e^{i \theta \globaldim (k - \bar{k})}  \targrandom{U(\mathcal{H})}{k, \bar{k}} \equiv \targrandom{U(\mathcal{H})}{k, \bar{k}}
\end{align}
As the equality should hold for any $\theta$, $\targrandom{U(\mathcal{H})}{k, \bar{k}} = 0$ if $k \neq \bar{k}$.

\subsection{Number-Conserving Unitary Group $U(\mathcal{H} | \hat{N})$}
\label{ssec:moments_ninv_unit_group}

With number conservation, the unitary group on the Hilbert space $\mathcal{H}$ is restricted to those unitaries which commute with the global number operator,
 \begin{align}
     U(\mathcal{H} | \hat{N}) &= \{ U \in U(\mathcal{H})\, |\, [U, \hat{N}] = 0\} \label{eq:nconservingugroup}
 \end{align}
 Any unitary in $U(\mathcal{H}\vert \hat{N})$ may be block decomposed
 \begin{align}
     U = \bigoplus_{n=0}^{M-1} U_n
     \label{Eq:NDecompofU}
 \end{align}
 where $U_n$ acts unitarily on $\mathcal{H}_n$ and is zero for all other sectors.
 The group $U(\mathcal{H}\vert\hat{N})$ is a compact subgroup of $U(\mathcal{H})$, isomorphic to the direct product of unitary groups on the $n$-subspaces.

The $U(\mathcal{H}\vert\hat{N})$ Haar ensemble is the unique shift-invariant ensemble on the number-conserving unitary group. 
In particular, if $U = \bigoplus_n U_n$ is sampled from the $U(\mathcal{H}\vert\hat{N})$ Haar ensemble, then each $U_n$ is independently sampled from the Haar ensemble on $U(\mathcal{H}_n)$, with $\mathcal{H}_n$ the Hilbert space of states with total number $n$. 
This allows us to bootstrap results in Sec.~\ref{ssec:moments_gen_unitary_group} to the number-conserving case.

Using Eq.~\eqref{Eq:NDecompofU}, the $k$-th moment operator is: 
\begin{align}
\label{eq:moment_of_ninv_U}
    \tkunhaar =  \unitaryaverage{ \bigoplus_{\vec{n}, \bar{\vec{n}}}\left( \bigotimes^k_{\alpha=1} U_{n_\alpha} \otimes U^*_{\bar{n}_\alpha}\right)}{U}
\end{align}
Here $\alpha$ labels the replicas and $\vec{n}$, $\bar{\vec{n}}$ labels the number sector for the replicas and $*$-replicas. 
Once again, shift- and inversion-invariance of the Haar measure imply that $\tkunhaar$ acts as a Hermitian projector on $\left(\mathcal{H} \otimes \mathcal{H}^*\right)^{\otimes k}$.
In addition to $S_k\times S_k$ replica symmetry, $\tkunhaar$ is invariant under $2k$ replicated number operators,
\begin{align}
\label{Eq:NumberReplica}
    [\tkunhaar, \hat{N}_\alpha]&=0 &
    [\tkunhaar, \hat{\bar{N}}_\alpha] &= 0 ,
\end{align}
which generate a $U(1)^k \times U(1)^k$ symmetry group.

Let us explicitly construct a set of states which span the $k$'th moment space for $U(\mathcal{H}|\hat{N})$. 
We begin with the parent permutation states $\ket{\sigma}$ of Eq.~\eqref{eq:sigmastates}; as $U(\mathcal{H}|\hat{N})$ is a  subgroup of $U(\mathcal{H})$, these states are automatically $+1$ eigenstates of the number-conserving moment operator, Eq.~\eqref{eq:moment_of_ninv_U}. 
The number symmetry, Eq.~\eqref{Eq:NumberReplica}, implies that we can act on $\ket{\sigma}$ with any operator constructed from $\hat{N}_\alpha$ and $\hat{\bar{N}}_\alpha$ and obtain another $+1$ eigenstate. 
In particular, the $U(1)^k$ \emph{phase states}, $\edgeket{\vec{\Phi}}{\sigma} = e^{i \sum_\alpha \Phi_\alpha \hat{N}_\alpha} \ket{\sigma}$, are all $+1$ eigenstates of $\tkunhaar$. 
We note that the permutation states are invariant under the $k$ generators $\hat{N}_\alpha - \hat{\bar{N}}_{\sigma(\alpha)}$, so that there are only $k$ phases available to parameterize the moment space.

Since they are continuously parameterized, the phase states are clearly linearly dependent. 
An alternative, discrete basis may be obtained by projecting $\ket{\sigma}$ onto the number sectors $\vec{n} = (n_1, \cdots, n_k)$ of each replica,
\begin{equation}
\begin{aligned}
    \edgeket{\vec{n}}{\sigma} &= \Pi_{\vec{n}} \ket{\sigma}\\
    &= \sum_{\vec{i}} \underbrace{\left(\prod_\alpha \delta[N_\alpha(i_\alpha)= n_\alpha] \right)}_{\delta[n(\vec{i})=\vec{n}]}
    \ket{\vec{i},\sigma(\vec{i})}
    \label{eq:numberstates}
\end{aligned}
\end{equation}
It is straightforward to check that, (i) states in Eq.~\eqref{eq:numberstates} with different number labels are orthogonal to one another, (ii) there are $M^k k!$ such states, (iii) the states are linearly independent if $k \le d_n$ for all $n$, and span the $+1$ eigenspace of $\tkunhaar$, and (iv) a linearly independent and spanning subset can be chosen if $k > d_n$ for some $n$. See Appendix~\ref{app:numberconservinggates}.
We thus obtain a Weingarten like representation of the $k$-th moment operator for $U(\mathcal{H} \vert \hat{N})$
\begin{equation}
    \begin{aligned}
        \tkunhaar &= \sum_{\vec{n}}\sum_{\sigma,\tau} \edgeket{\vec{n}}{\sigma} V_{\vec{n}}^{\sigma\tau} \edgebra{\vec{n}}{\tau}.\label{eq:ninvariantaverage}
    \end{aligned}
\end{equation}

As in the non-conserving case in Sec.~\ref{ssec:moments_gen_unitary_group}, the coefficients $V_{\vec{n}}^{\sigma\tau} $ follow from the inverse of a suitably restricted overlap matrix for any value of $k$. Explicit formulae follow from a straightforward generalization of Appendix~\ref{App:Vsigmatau}.

We note that the states $\edgeket{\vec{\Phi}}{\sigma}$ transform non-trivially under the $U(1)^k \times U(1)^k$ number symmetry and the $S_k \times S_k$ replica symmetry. In the language of the ground state space of the Hamiltonian $\hat{H}_k = 1-\tkunhaar$, we say that the $U(1)^k \times U(1)^k$ number symmetry is spontaneously broken down to the diagonal subgroup $U(1)^k$ which is intertwined with the diagonal $S_k$ chosen by the $S_k \times S_k$ replica symmetry breaking.

The unbalanced moment operators, number sector by number sector, are identically zero following the same argument as in Sec.~\ref{ssec:moments_gen_unitary_group}.

\section{Moments of $b$-body Circuit Ensembles}
\label{sec:moments_ng_unit_group}

We now turn our attention to the group of unitaries generated by $b$-body circuits -- that is, circuits of arbitrary depth composed of unitary gates which act on at most $b$ qudits. 
Without number conservation, this group is simply the full unitary group, $U_b(\mathcal{H}) = U(\mathcal{H})$~\cite{MikeAndIke}.
As noted in Eq.~\eqref{eq:conv_T_haar}, this implies that the moment operators of brick-layer random circuits converge to the moment operators of the full unitary group.

It came as a surprise when Marvian~\cite{marvian_restrictions_2022} showed that number-conserving $b$-body circuits fail to generate the full group of number-conserving unitaries, $U(\mathcal{H}\vert\hat{N})$.
Rather, $b$-body circuits generate a proper subgroup $U_b(\mathcal{H}\vert\hat{N}) \subset U(\mathcal{H}\vert\hat{N})$.
However, since the codimension of $U_b(\mathcal{H}\vert\hat{N})$ is very small (linear in $L^d$) compared to the full dimension of  $U(\mathcal{H}\vert\hat{N})$ (exponential in $L^d$), one might expect that the moment operators are distinguishable only for rather high-order moments.
Indeed, we argue here that this is the case: in the thermodynamic limit $L\to\infty$, \emph{all} finite moments of the $b$-body circuit ensemble converge to those of the Haar measure on $U(\mathcal{H}\vert\hat{N})$. 
More precisely, the moment index $k$ must exceed a critical value $k_{c}\geq L^d$ to distinguish between the Haar measure on $U(\mathcal{H}\vert\hat{N})$ and the Haar measure on $U_{b}(\mathcal{H}\vert\hat{N})$.

When restricted to number sector $n$, elements of $U_b(\mathcal{H}|\hat{N})$ and $U(\mathcal{H}|\hat{N})$ differ by an overall phase~\cite{marvian_restrictions_2022}.
In Sec.~\ref{subs:decomp_moment_op}, we show that $\tkunhaar$ and $\tkughaar$ can each be factored into the product of two moment operators: a projector on the space of $U(1)$ charges and $\tkrandom{\text{SU}(\mathcal{H}\vert\hat{N})}$ for a related group of special unitary matrices $\text{SU}(\mathcal{H}\vert\hat{N})\equiv\prod_n^M \text{SU}(\mathcal{H}_n)$.
The operators $\tkunhaar$ and $\tkughaar$ differ only in the first factor, see Eq.~\eqref{eq:MomentDifference}.
In Sec.~\ref{subs:first_factor}, we impose shift-invariance of the Haar measure to obtain linear constraints, Eq.~\eqref{eq:PhaseConstraints}, on the number sectors in which the difference can be nonzero. 
In Sec.~\ref{subs:second_factor}, we turn to the second factor, $\tkrandom{\text{SU}(\mathcal{H}\vert\hat{N})}$, which provides further constraints on the number sectors in which the difference is nonzero, Eq.~\eqref{eq:su_number_constraint}.
Finally, in Sec.~\ref{subs:boundonkc}, we combine these constraints to obtain $k_{c} \ge L^d$.

\subsection{Decomposition of Moment Operators}
\label{subs:decomp_moment_op}
Consider a number-conserving unitary $U \in U(\mathcal{H}\vert\hat{N})$. 
It may be block diagonalized,
\begin{align}
\label{eq:ndecompu_factorization}
    U& =\bigoplus_{n}e^{i\theta_n}\tilde{U}_{n},
\end{align}
where $\det\tilde{U}_{n}=1$. 
This factorization is not unique as one may shift $\theta_n \to \theta_n + 2\pi / d_n$ and simultaneously multiply $\tilde{U}_n$ by a compensating phase.
Nonetheless, the Haar measure on $U(\mathcal{H}\vert\hat{N})$ factorizes uniformly,
\begin{equation}
    \int dU = \int \left(\prod_n^{M} \frac{d\theta_n}{2\pi} d\tilde{U}_n\right) = 1
\end{equation}
From these factorizations, 
\begin{widetext}
\begin{align}
\label{eq:HaarDecomposition}
    \tkunhaar &= \int dU\left(\bigoplus_{n}e^{i\theta_{n}}\tilde{U}_{n}\otimes\bigoplus_{\bar{n}}e^{-i\theta_{\bar{n}}}\tilde{U}^{*}_{\bar{n}}\right)^{\otimes k} \nonumber \\
    &= \bigoplus_{\{n_\alpha, \bar{n}_\alpha\}} 
     \underbrace{\int \left(\prod_n^M \frac{d\theta_n}{2\pi}\right) \exp{i\sum_{\alpha}\left(\theta_{n_\alpha}- \theta_{\bar{n}_\alpha} \right)}}_{\hat{P}_{\vec{n}\bar{\vec{n}}}}
     \times
     \underbrace{\int \left(\prod_n^M d\tilde{U}_n\right) \bigotimes_\alpha^k \left(\tilde{U}_{n_\alpha} \otimes \tilde{U}_{\bar{n}_\alpha}^* \right)}_{(\tkrandom{\text{SU}(\mathcal{H}\vert\hat{N})})_{{\vec{n}\bar{\vec{n}}}}} \nonumber\\
     &= \hat{P} ~ \tkrandom{\text{SU}(\mathcal{H}\vert\hat{N})} 
\end{align}
\end{widetext}
In the final step, we have exploited that both $\hat{P}$ and $\tkrandom{\text{SU}(\mathcal{H}\vert\hat{N})}$ are block diagonal in the $M^{2k}$ number blocks labeled by $\vec{n}, \bar{\vec{n}}$.
Indeed, both $\hat{P}$ and $\tkrandom{\text{SU}(\mathcal{H}\vert\hat{N})}$ are moment operators in the sense of Sec.~\ref{ssec:generalproperties} for Haar measures on $U(1)^M$ and $SU(\mathcal{H}\vert\hat{N})$, respectively.

Let us now derive an analogous factorization to Eq.~\eqref{eq:HaarDecomposition} for the $b$-body unitary circuits, $U_b(\mathcal{H}|\hat{N})$. 
As demonstrated in Ref.~\onlinecite{marvian_restrictions_2022}, the structure of $b$-body unitary circuits imposes a set of linear constraints among the phases $\{\theta_{n}\}$ in~\eqref{eq:ndecompu_factorization}. 
In other words, although geometrically local number conserving circuits may realize any special unitary $\tilde{U} \in \text{SU}(\mathcal{H}\vert\hat{N})$, they may not realize arbitrary unitaries $U \in U(\mathcal{H}\vert\hat{N})$.
Turning back to the factorization~\eqref{eq:HaarDecomposition}, the constraints on $\{\theta_n\}$ may modify $\hat{P}$ but leave $\tkrandom{\text{SU}(\mathcal{H}\vert\hat{N})}$ unchanged.
We denote by $\{\phi_{\ell}\}$ the set of unconstrained phases which parameterize the $(b+1)$-dimensional subtorus of $U(1)^{M}$ accessible by $b$-body circuits. 
The moment operator $\tkughaar$ can then be written as
\begin{equation}\label{eq:GbodyDecomposition}
    \tkughaar = \hat{P}_b \tkrandom{\text{SU}(\mathcal{H}\vert\hat{N})} 
\end{equation}
where 
\begin{align}
\label{eq:pgdef}
    \hat{P}_b &= \bigoplus_{\{n_\alpha, \bar{n}_\alpha\}}  \int \left(\prod_{\ell}^{b+1} \frac{d\phi_{\ell}}{2\pi}\right)  \exp{i \sum_\alpha \left(\theta_{n_\alpha}-\theta_{\bar{n}_\alpha}\right)}
\end{align}
The constrained number sector phases $\{\theta_n\}$ now depend on the unconstrained phases $\{\phi_l\}$,
\begin{align}
    \theta_n = \sum_\ell G_{n\ell} \phi_\ell .
\end{align}
Here, the integer-valued matrix $G$ encodes the linear embedding of the $\phi$-torus into the $\theta$-torus. 
Explicit forms for $G$ are available in Ref.~\cite{marvian_restrictions_2022}, but are unnecessary for our purposes: what is essential is that $\hat{P}_{b}$ is a projector. 
This follows either by explicit evaluation of the Fourier representation Eq.~\eqref{eq:pgdef} or from  regarding $\hat{P}_{b}$ as a moment operator of the Haar measure on the $\phi$-torus $U(1)^{b+1}$.

Thus, the difference of the moment operators can be written,
\begin{equation}\label{eq:MomentDifference}
    \tkughaar - \tkunhaar = \left(\hat{P}_b - \hat{P}\right) \tkrandom{\text{SU}(\mathcal{H}\vert\hat{N})} 
\end{equation}
As is clear from the definitions in  Eqs.~\eqref{eq:HaarDecomposition},~\eqref{eq:pgdef}, the  factor, $\hat{P}_b - \hat{P}$, only depends on the number sector $\{n_\alpha, \bar{n}_\alpha\}$ and, as we show in Sec.~\ref{subs:first_factor}, vanishes except in special circumstances.
In those circumstances, we turn to the second factor, $\tkrandom{\text{SU}(\mathcal{H}\vert\hat{N})}$, to determine the minimal $k$ at which the moment operators are distinguishable, Sec.~\ref{subs:second_factor}.

\subsection{First Factor: Number Sector Projectors}
\label{subs:first_factor}

The projectors, $\hat{P}_b$ and $\hat{P}$, select number blocks in the replicated space. 
Without the $b$-body restriction, the projector $\hat{P}$ simply projects  onto number-balanced blocks. 
To make this more precise, define the counting operators,
\begin{align}
    \hat{\#}_n &= \sum_{\alpha} \delta[\hat{N}_\alpha = n] & 
    \hat{\bar{\#}}_n &= \sum_{\alpha} \delta[\hat{\bar{N}}_\alpha = n]
\end{align}
which determine the number of replicas ($*$-replicas) in the number sector $n$. 
As these operators are diagonal in number and in our choice of standard basis, Eq.~\eqref{eq:standardbasis}, we will also abuse notation slightly by writing $\#_n(\vec{n})$ and $\#_n(\vec{j})$ for the appropriate diagonal matrix elements of the operator $\hat{\#}_n$ (and similarly for $\hat{\bar{\#}}_n$). 
The Fourier integral defining $\hat{P}$, Eq.~\eqref{eq:HaarDecomposition}, imposes that these counts cancel,
\begin{align}
    \hat{P} = \prod_{n}^M \delta[\hat{\#}_n = \hat{\bar{\#}}_n]
\end{align}
as claimed.

To understand the projector $\hat{P}_b$, we make use of the shift invariance of the Haar measure on $U(1)^{b+1}$. 
Consider the effect of a global 0-body phase shift $\gamma$, which acts on operators $U\in U(\mathcal{H}|\hat{N})$ as $U\to e^{i\gamma}U$. 
This transformation acts entirely on the phase part of the factorization, Eq.~\eqref{eq:ndecompu_factorization}, which leads  matrix elements of $\hat{P}_{b}$ to acquire non-trivial phases,
\begin{align}\label{eq:GlobalPhaseShift}
\bra{\vec{i},\bar{\vec{i}}}\hat{P}_{b}\ket{\vec{i},\bar{\vec{i}}} \to \exp\left[i \gamma \sum_n \left(\#_n(\vec{i}) - \bar\#_n(\bar{\vec{i}})\right)\right] \nonumber \\ \hfill \times  \bra{\vec{i},\bar{\vec{i}}}\hat{P}_b\ket{\vec{i},\bar{\vec{i}}}
\end{align}
Similarly, we can implement 1-body phase shifts $U = \prod_{x,\alpha} e^{i \beta \hat{n}_{x,\alpha}}$, under which matrix elements transform as
\begin{align}\label{eq:1BodyShifts}
\bra{\vec{i},\bar{\vec{i}}}\hat{P}_b\ket{\vec{i},\bar{\vec{i}}} \to \exp\left[i\beta\sum_{n}n\left(\#_n(\vec{i}) - \bar\#_n(\bar{\vec{i}})\right)\right]\nonumber \\ \hfill \times \bra{\vec{i},\bar{\vec{i}}}\hat{P}_b\ket{\vec{i},\bar{\vec{i}}}
\end{align}
However, shift invariance of the Haar measure guarantees that both global phase shifts and 1-body phase shifts leave $\hat{P}_{b}$ unchanged (for $b \ge 1$); equations~\eqref{eq:GlobalPhaseShift} and~\eqref{eq:1BodyShifts} are therefore satisfied for arbitrary $\gamma, \beta$ only if
\begin{equation}\label{eq:PhaseConstraints}
\begin{aligned}
\sum_{n}\left(\#_n - \bar{\#}_n\right)&=0\\
\sum_{n}n\left(\#_n - \bar{\#}_n\right)&=0.
\end{aligned}
\end{equation} 
The number balanced sectors, or the $+1$ eigenspace of $\hat{P}$, provide a trivial solution of these constraints, thus $\hat{P}_b$ projects onto a superspace of the image of $\hat{P}$, as expected.

It follows that the moment operators, $\tkughaar$ and $\tkunhaar$, are indistinguishable within number-balanced blocks of the replicated space. 
Thus, in order to find differences, we need to consider nontrivial solutions to the constraints Eq.~\eqref{eq:PhaseConstraints}.
We return to these constraints to construct a bound on the critical moment index, $k_{c}$, in Sec.~\ref{subs:boundonkc}.

\subsection{Second Factor: Moments of the Special Unitary Group}
\label{subs:second_factor}

Consider the moment operator $\tkrandom{\text{SU}(\mathcal{H}\vert\hat{N})}$ for the Haar measure on the special unitary group $SU(\mathcal{H}|\hat{N})\equiv\prod_n^M SU(\mathcal{H}_n)$.
The operator is block-diagonal with $M^{2k}$ blocks, labeled by the replicated numbers, $\vec{n}$, and $\bar{\vec{n}}$.
Here, we show that if the block $(\tkrandom{\text{SU}(\mathcal{H}\vert\hat{N})})_{\vec{n}\bar{\vec{n}}}$ is non-zero, then $\#_n(\vec{n}) - \bar{\#}_n(\vec{\bar{n}})$ is an integer multiple of the dimension $d_n$ for every $n=0,\cdots, M-1$.

Consider the unitary $W \in SU(\mathcal{H}|\hat{N})$
\begin{align}
    W &= \exp\left({i \frac{2\pi}{d_n} \delta[\hat{N} = n]}\right)
\end{align}
which acts as the identity in all number sectors except $n$, in which it applies a global phase $e^{i 2\pi/d_n}$. 
Right shift invariance of the Haar measure implies,

\begin{equation}\label{eq:shift_dn}
\begin{aligned}
          \exp\left[i \frac{2\pi}{d_n} \left( \#_n(\vec{j}) - \bar\#_n(\vec{\bar{j}}) \right) \right]& 
          \tkrandom{\text{SU}(\mathcal{H}\vert\hat{N})}  \ket{\vec{j},\bar{\vec{j}}}\\ 
        &\hspace{10pt}=  \tkrandom{\text{SU}(\mathcal{H}\vert\hat{N})} \ket{\vec{j},\bar{\vec{j}}}  .
\end{aligned}
\end{equation}
If $\tkrandom{\text{SU}(\mathcal{H}\vert\hat{N})} \ket{\vec{j},\bar{\vec{j}}}$ is nonzero with $\vec{j}$, $\bar{\vec{j}}$ in the $\vec{n}$, $\bar{\vec{n}}$ number sector, then we must have that the phase in Eq.~\eqref{eq:shift_dn} is 1. Therefore,
\begin{align}
\label{eq:su_number_constraint}
\#_n(\vec{n}) - \bar\#_n(\vec{\bar{n}})  = c_n d_n
\end{align}
for some integer $c_n$.

\subsection{A Bound on $k_{c}$}
\label{subs:boundonkc}

According to Sec.~\ref{subs:first_factor}, in order to distinguish $\tkughaar$ from $\tkunhaar$, we must look in an unbalanced number block where $\#_n \neq \bar\#_n$ for some $n$; this implies that the constraint Eq.~\eqref{eq:PhaseConstraints} holds non-trivially.
Such blocks vanish in the common factor $\tkrandom{\text{SU}(\mathcal{H}\vert\hat{N})}$ unless they further satisfy Eq.~\eqref{eq:su_number_constraint}. 
Combining these two conditions, we find the linear system
\begin{equation}\label{eq:PhaseConstraintsRewrite}
\begin{aligned}
\sum_{n}  d_{n} c_{n}&=0\\
\sum_{n}n d_{n} c_{n}&=0
\end{aligned}
\end{equation} 
where $c_n$ are the integers in Eq.~\eqref{eq:su_number_constraint}.

Any nontrivial vector of integers $\vec{c} = (c_0 \cdots c_{M-1})$ which satisfy the conditions~\eqref{eq:PhaseConstraintsRewrite} must contain a minimum of three nonzero entries. 
With only two, the restricted coefficient matrix always has a full rank and only admits zero solutions.  
Translating back to $\vec{n}, \vec{\bar{n}}$, there must be at least three unbalanced number sectors.

Due to Eq.~\eqref{eq:su_number_constraint}, any unbalanced number sector $n$ has $|\#_n(\vec{n}) - \bar\#_n(\vec{\bar{n}})| = |c_n d_n| \ge d_n$. 
This requires $k \ge d_n$. 
Accordingly, the minimum value of $k$ to distinguish $\tkughaar$ from $\tkunhaar$ grows at least as quickly as the dimension of the third-smallest number sector. 
A worst-case estimate for the scaling of different number sectors is provided by qubit systems with $S_{z}$ conservation, defined by single-particle dimensions $d^{(1)}=(1,1)$.
The two smallest number sectors correspond to polarized states, $d_{0}=d_{V}=1$. 
The third smallest sector, which is degenerate with $d_{1}=d_{V-1}=V$, grows extensively, corresponding to the number of choices of qubit flips in the fully polarized states. We therefore conclude that 
\begin{align}
    k_{c}\geq V
\end{align} 
in qubit systems. 

Any other qudit structure, $d^{(1)}$, leads to at least as large a lower bound on $k_c$, typically exponentially large in $V$.

\section{Gap Analysis of Circuit Moment Operators}
\label{sec:circuits}

Now we turn to an analysis of the gap of the quantum statistical mechanical system defined by replica averaging the random circuits of depth $t$, see Fig.~\ref{fig:random_circuit}. 
This gap controls the convergence of the finite depth moments, $\tkrandom{U_t}$, to the global Haar moments, $\tkunhaar$.
Throughout this section, we assume $k < k_c$ and neglect the distinction between $U(\mathcal{H}|\hat{N})$ and $U_b(\mathcal{H}|\hat{N})$.

We analyze two closely related statistical models in $d+1$ dimensions. 
The primary model of interest gives the imaginary time representation of the moment operator, with partition function,
\begin{align}
    Z^{(k)}_t &= \int dU~p(U_t=U) \vert\Tr U\vert^{2k} = \Tr[\tkrandom{U_t}].
    \label{eq:repcircpartition}
\end{align}
The associated transfer matrix,
\begin{align}
    \tkrandom{U_l} = \prod_{\text{Gates }g \in l} \tkrandom{U_g} 
\end{align}
propagates the system between layers in imaginary time.
This model spontaneously breaks the $U(1)^k\times U(1)^k$ symmetry down to $U(1)^k$ in a frustration-free manner. 
In Sec.~\ref{ssec:circuit_gap}, we bound the gap of the transfer matrix due to the associated Goldstone modes.

Before turning to the circuit model, in Sec.~\ref{ssec:VariationalHam} we analyze the closely related, but technically simpler, Hamiltonian system
\begin{align}
    \hat{H}_k &= \sum_{\text{Gates~}g \in l} (1 - \tkrandom{U_g})
    \label{Eq:HamFF2}
\end{align}
where the sum runs over all gates present in a layer of the replicated circuit.
The ground state space of $\hat{H}_k$ is constructed to be identical to the $+1$ eigenspace of the layer transfer matrix $\tkrandom{U_l}$. 
Again, the Goldstone modes upper bound the Hamiltonian gap, which in turn controls the large imaginary time properties.

Both models have gaps scaling as $O(1/L^{2})$. 
This is the result we use in Sec. \ref{sec:kdesignconvergencetime} to bound the convergence of the circuit to an $\varepsilon$-approximate $k$-design.

\subsection{Variational Bound on Gap of $\hat{H}_k$}
\label{ssec:VariationalHam}

As discussed in Sec.~\ref{sec:intro}, the ground states of $\hat{H}_k$ are spanned by the phase states, 
\begin{align}
    \edgeket{\vec{\Phi}}{\sigma} &= \edgeket{\Phi_1,\ldots,\Phi_k}{\sigma} = e^{i\sum_\alpha \Phi_\alpha \hat{N}_\alpha } \ket{\sigma}
\end{align}
These states transform non-trivially under $U(1)$ rotations associated with the conserved charges $\hat{N}_{\alpha}$, and $\hat{\bar{N}}_\alpha$,
\begin{align}
    e^{i\left(\theta \hat{N}_{\alpha} +\bar\theta \hat{\bar{N}}_{\sigma(\alpha)}\right)} \edgeket{\vec{\Phi}}{\sigma}=\edgeket{\Phi_1,\ldots,\Phi_\alpha  + \theta + \bar\theta,\ldots,\Phi_k}{\sigma}
\end{align}
Notably, for any choice of $\theta$, the ground state is invariant under $e^{i \theta (\hat{N}_{\alpha} - \hat{\bar{N}}_{\sigma(\alpha)})}$ but transforms nontrivially under $e^{i \theta (\hat{N}_{\alpha} + \hat{\bar{N}}_{\sigma(\alpha)})}$.
Thus, of the original $2k$ symmetries, $k$ are unbroken.

Each broken symmetry yields a Goldstone mode.
These excited states closely resemble ground states $\edgeket{\Phi(x)}{\sigma}$ in a small region around a given position $x$.
If $\Phi(x)$ varies slowly with $x$ we may expect the excitation energy, which is a sum of local terms, to be small.
To upper bound the gap, $\Delta$, we construct a low energy state with long-wavelength modulations of the phase, 
\begin{align}
    e^{i \theta \sum_x e^{-iqx} \hat{n}^\alpha_x} \ket{\Psi} &= (1 + i \theta \hat{n}^\alpha_q + ...) \ket{\Psi} \nonumber\\&= \ket{\Psi} + i \theta \ket{q_\alpha} + ...
\end{align}
where $\hat{n}_q^\alpha = \sum_x e^{-iqx} \hat{n}_x^\alpha$ and $\ket{\Psi}$ is an arbitrary ground state of $\hat{H}_k$ and $\theta$ is a perturbative parameter.
The finite momentum states $\ket{q_\alpha}$ are necessarily orthogonal to the ground states of $\hat{H}_k$,
which have leg-permutation symmetry (see Eq.~\eqref{eq:tlegsym}) and therefore carry zero momentum.
The variational energy 
\begin{align}
    \Delta_q \equiv \frac{\bra{q_\alpha} \hat{H}_k \ket{q_\alpha}}{\braket{q_\alpha}{q_\alpha}} . \label{eq:gapdefinition}
\end{align}
upper bounds the true gap $\Delta$.
Note that the replica index, $\alpha$, is just a spectator and is omitted going forward. 

We now evaluate the numerator of \eqref{eq:gapdefinition} using symmetries of $\hat{H}_k$,
\begin{align}
    \bra{q} \hat{H}_k \ket{q} &= \sum_{x,y,g} e^{iq(y-x)} \langle \hat{n}_x (1 - \tkrandom{U_g}) \hat{n}_y \rangle . \label{eq:fouriersum}
\end{align}
Here and henceforth, $\langle \cdot \rangle$ implicitly denotes an average with respect to the ground state $\ket{\Psi}$. We further specialize to the case of $2$-body gates in a nearest neighbor square lattice for simplicity, although the final result can be readily generalized.
Each gate operator $\tkrandom{U_g}$ commutes with the total number operator for sites in the support of that gate. 
For example, if gate $g$ acts on $x$ and $y$,
\begin{align}
    [\tkrandom{U_{xy}}, \hat{n}_x + \hat{n}_y] = 0 .
\end{align}
Further, leg permutation symmetry, Eq.~\eqref{eq:tlegsym}, and $\tkrandom{U_{xy}} \ket{\Psi} = \ket{\Psi}$ implies an identity for number operators connected by a gate, 
\begin{align}
    \tkrandom{U_{xy}} \hat{n}_x \ket{\Psi} &= \frac{1}{2} (\hat{n}_x + \hat{n}_y) \ket{\Psi} .\label{eq:naveragingidentity}
\end{align}

Applying \eqref{eq:naveragingidentity} to each term in \eqref{eq:fouriersum}, we remove all dependence on the moment operator and replica count $k$,
\begin{align}\label{eq:numerator}
    \langle \hat{n}_x \tkrandom{U_{xy}} \hat{n}_x \rangle = \langle \hat{n}_x \tkrandom{U_{xy}} \hat{n}_y \rangle &= \frac{\langle \hat{n}_x^2 \rangle + \langle \hat{n}_x \hat{n}_y \rangle}{2} .
\end{align}
Correlation functions of number operators are strongly constrained by invariance under translations and permutations:
\begin{equation}\label{eq:NumberCorrelatorSymmetry}
    \langle\hat{n}_{x}\hat{n}_{y}\rangle = \langle\hat{n}_{0}\hat{n}_{y-x}\rangle = \delta_{x,y}\langle\hat{n}_{0}^{2}\rangle+\left(1-\delta_{x,y}\right)\langle\hat{n}_{0}\hat{n}_{1}\rangle
\end{equation}
where in the first equality we have applied translation invariance, and in the second, permutation invariance. Applying this result to Eq.~\eqref{eq:numerator}, we find
\begin{align}
    \bra{q} \hat{H}_k \ket{q} &= L^d \left(\langle \hat{n}_0^2 \rangle - \langle \hat{n}_0 \hat{n}_1 \rangle\right)\left( d - \sum^d_\mu \cos q_\mu \right) . \label{eq:numersimple}
\end{align}

We now turn our attention to the denominator of \eqref{eq:gapdefinition}. Using the structure of correlation fuctions in Eq.~\eqref{eq:NumberCorrelatorSymmetry}, we find
\begin{align}
    \braket{q}{q} &= L^d \langle \hat{n}_0^2 \rangle + L^d \langle \hat{n}_0 \hat{n}_1 \rangle (\delta_{q,0} L^d - 1) . \label{eq:denomsimple}
\end{align}
Putting \eqref{eq:numersimple} and \eqref{eq:denomsimple} together we find a variational gap,
\begin{align}
    \Delta_q &= d - \sum_\mu \cos q_\mu . \label{eq:hamiltoniangap}
\end{align}
Note that $\Delta_q$ is independent of our initial choice of $\ket{\Psi}$ as well as the qudit structure.
One exception, however, are fully polarized states, which are eigenstates of the number operator on each site and have no number fluctuations; in these sectors, one cannot construct a state with nonzero momentum and the variational analysis breaks down.
However, it is clear that in such cases that conservation laws are irrelevant, and the results of previous studies\cite{hunter-jones2019,jian2022} predict an $O(1)$ gap.

All finite momentum states are orthogonal to the ground state manifold and the variational gap $\Delta_q$ provides an upper bound for $\Delta$.
Minimization of $\Delta_q$ with respect to $q$ in a finite system provides the bound on the gap in the square lattice system
\begin{align}
    \Delta \le \frac{2\pi^2}{L^2}.
\end{align}

Careful readers may have recognized that Eq.~\eqref{eq:hamiltoniangap} is the spatial Fourier transform of the Laplace operator on the square lattice. 
This is no accident; on general lattices with higher body interactions, the charge undergoes a continuous time random walk on the spatial lattice.
Consider the imaginary time correlation function (for $t>0$),
\begin{align}
    C_{xy}(t) &= \langle \hat{n}_x(t) \hat{n}_y(0) \rangle 
\end{align}
where
\begin{align}
    \hat{n}_x(t) &= e^{+t H} \hat{n}_x e^{-t H}
\end{align}
The correlator satisfies the equation of motion,
\begin{align}\label{eq:ctrw_eom}
    \frac{d}{dt} C_{xy}(t) &= \langle [H, n_x(t)] n_y \rangle = \langle \hat{n}_x(t) H \hat{n}_y \rangle 
\end{align}
where in the second equality we have used the frustration-free condition. 
That is, the right hand side is precisely the numerator of Eq.~\eqref{eq:gapdefinition}.
Using number conservation and leg permutation symmetries of the gates it is straightforward to show that $[H, \hat{n}_x]$ is a linear combination of $\hat{n}_{x'}$ operators when evaluated within the ground space. 
This leads Eq.~\eqref{eq:ctrw_eom} to close,
\begin{align}
    \frac{d}{dt} C_{xy}(t) &= - \sum_{x'} \Gamma_{x',x} C_{x'y}(t)
\end{align}
where $\Gamma_{x', x}$ is a transition rate matrix which satisfies $\sum_{x'} \Gamma_{x', x} = 0$ in order to conserve number.
That is, we have obtained a continuous time random walk.
For any particular choice of gate geometry, it is straightforward to compute $\Gamma$ explicitly. 

For a general short-ranged, translation-invariant Hamiltonian, we can Fourier transform $\Gamma$ and find a long-wavelength gap which closes as $q^2$. Thus, we obtain
\begin{align}
    \Delta \le \frac{C}{L^2}.
\end{align}
where $C$ is a geometry dependent factor.

\subsection{Variational Gap on $\tkrandom{U_t}$}
\label{ssec:circuit_gap}

We now consider the layer transfer matrix, $\tkrandom{U_l}$, governing the primary circuit model.
There are two technical complications compared to the Hamiltonian case.
First, as $\tkrandom{U_l}$ is not Hermitian, it is not guaranteed to have real eigenvalues. 
However, it is frustration-free and has a degenerate $+1$ eigenspace identical to the ground state space of the Hamiltonian.
The relevant gap is thus between the $+1$ eigenvalues and the real part of the next-largest eigenvalue. 
Second, the circuit $\tkrandom{U_{l}}$ has structure in time corresponding to the sequence in which gates are applied.
This complicates the mathematical treatment slightly, but does not change the scaling of the Goldstone gap with $L$.

Recall that the moment operator for depth $t$ circuits can be broken into $t$ layers composed of individual gates obtained by Haar-averaging,
\begin{align}
    \tkrandom{U_t} &= \left(\tkrandom{U_l} \right)^t .
\end{align}
We write the second-largest-magnitude eigenvalue of $\tkrandom{U_l}$ as $\lambda \equiv e^{-\Delta}$, with $\text{Re}(\Delta) > 0$.  
To bound $\Delta$, we again consider the Goldstone excitations, $\ket{q} = \hat{n}_q \ket{\Psi}$.
The reduction in norm of $\ket{q}$ under the action of the transfer matrix sets a bound,
\begin{align}
    \frac{\vert\vert \tkrandom{U_l} \ket{q} \vert\vert^2}{\braket{q}{q}} \le \vert e^{- 2 \Delta} \vert . \label{eq:circuitboundfractioncomplex}
\end{align}
The denominator is familiar -- see Eq.~\eqref{eq:denomsimple}.

The numerator of Eq.~\eqref{eq:circuitboundfractioncomplex} can be mapped onto a discrete time random walk reminiscent of the continuous time walk obtained in the Hamiltonian case.
Every step of the walk spreads the operators $\hat{n}_{x}$ equally to the outgoing legs of each gate, so that the moves of the conserved charge are determined by the geometry of the circuit. 
This process is encoded in a doubly-stochastic matrix $R$,
\begin{align}
    \tkrandom{U_l} \hat{n}_x \ket{\Psi} \equiv \sum_y R_{yx} \hat{n}_y \ket{\Psi} . \label{eq:randomwalkdef}
\end{align}
Following similar arguments used to simplify the Hamiltonian bound, Eq.~\eqref{eq:circuitboundfractioncomplex} can be rewritten as
\begin{align}
    \frac{1}{L^d} \sum_{xya} R_{xa} R_{ay} e^{iq(x-y)} \le \vert e^{- 2 \Delta} \vert \label{eq:rtobound}.
\end{align}
It is useful to define $\tilde{R}_{xy} \equiv \sum_a R_{xa} R_{ay}$, which is itself a symmetric, doubly-stochastic matrix corresponding to the walk on the circuit defined by $(\tkrandom{U_l})^\dagger \tkrandom{U_l}$.
For small $q$ and sufficiently short-ranged $\tilde{R}$, 
\begin{widetext}
\begin{align}\label{eq:RandomWalkBound}
    \frac{1}{L^d} \sum_{xy} \tilde{R}_{xy} e^{iq(x-y)} &= \frac{1}{L^d} \sum_{xy} \tilde{R}_{xy} \cos\left(q(x-y)\right)\nonumber \\
    &\simeq 1 -  q^2 \underbrace{\left(\frac{1}{2 L^d}\sum_{xy} (x-y)^2 \tilde{R}_{xy}\right)}_{\tilde{C}} + ~O(q^4)
\end{align}
\end{widetext}
where $\tilde{C}$ is a geometry dependent constant.
Plugging this into Eq.~\eqref{eq:rtobound} and selecting the longest wavelength available, leads to 
\begin{align}
    \text{Re}(\Delta) \le \frac{C}{L^2} \label{eq:deltageometric}
\end{align}
for another constant $C$.
In Appendix~\ref{app:tl1dbrick}, we explicitly compute the bound including constants for the case of 2-body brick-layer circuits in $d=1$ and obtain a bound $\frac{4\pi^2}{L^2} \ge \Delta$.

\subsection{Lower Bounds on the Spectral Gap}

We conclude this section by briefly discussing the problem of lower bounding the spectral gap of $\hat{H}_k$.
So far, we have constructed variational states which upper bound the spectral gap by $O(L^{-2})$ for circuits in any number of spatial dimensions.
These bounds agree with the existing upper bounds for general frustration free translationally invariant Hamiltonians in one and two spatial dimensions~\cite{Gosset_Mozgunov_2016}, and, more recently, in higher dimensions~\cite{Masaoka_Soejima_Watanabe_2024} and general graphs with finite range interactions~\cite{Lemm_Lucia_2024}.
In the next section we show that this upper bound on the gap can be used to rigorously lower bound the convergence time to approximate $k$-designs.
In other words, we show that there is a minimum circuit depth before which the random circuits cannot form approximate designs.
In order to \emph{upper} bound the convergence time, and in effect guarantee convergence for sufficiently deep circuits, we need a rigorous lower bound on the spectral gap of $\hat{H}_k$ or $\tkrandom{U_t}$.

A large body of work focuses on the problem of lower bounding the spectral gap of frustration free translationally invariant Hamiltonians.
Broadly, there are two methods in one and two spatial dimensions~\cite{Knabe_1988,Nachtergaele_1996}.
Both methods consider local relations between the ground state space and the lowest lying excitations as additional sites are added to the system.
These methods are best equipped to bound gapped Hamiltonians, and can be applied alongside known upper bounds on convergence depths to bound the spectral gap of circuits without number-conservation~\cite{brandao2016b}.
In the case of number-conserving circuits, however, the spectral gap vanishes in the large $L$ limit and both techniques fail to lower bound the gap.
We thus leave obtaining a rigorous lower bound on the gap of the number-conserving circuits as an interesting open question.

\section{Convergence Time to Approximate Unitary $k$-Designs}
\label{sec:kdesignconvergencetime}

The variational upper bound on the spectral gap due to the Goldstone analysis of Sec.~\ref{sec:circuits} leads to a rigorous lower bound on the time $\tau$ at which the circuit ensemble becomes an approximate $k$-design.
If we further conjecture that there are no lower energy states which the Goldstone analysis misses, we obtain an upper bound on $\tau$ as well. 
In this section, we derive these lower and upper bounds on $\tau$.

The circuit ensemble is said to be an $\varepsilon$-approximate $k$-design when the diamond distance satisfies
\begin{align} \label{eq:convergencebound}
    \norm{\tkrandom{U_t} - \tkunhaar}_\diamond \le \varepsilon .
\end{align}
Here, $ \vert\vert\cdot \vert\vert_{\diamond}$ denotes the diamond norm~\cite{aharonov1998quantum}, and $\varepsilon$ the desired \emph{additive} error tolerance.
The bound in Eq.~\eqref{eq:convergencebound} can be interpreted as a bound on the probability of successfully distinguishing the circuit ensemble from the Haar ensemble using a single measurement~\cite{Low:2010aa}.
The diamond norm is bounded from above and below by the simpler $2$-norm \cite{hunter-jones2019,watrous_2018},
\begin{align}
    \norm{A}^2_2 \le \norm{A}^2_\diamond \le \globaldim^{2k} \norm{A}^2_2 \label{eq:inequalities}
\end{align}
It is therefore sufficient to estimate the 2-norm distance to establish both a lower and upper bound on $\tau$.

To proceed, we note that the moment operator admits an eigendecomposition
\begin{align}
    \tkrandom{U_t} =  \tkunhaar +  \sum_{i}\hat{C}_i e^{-t\Delta_i} \label{eq:eigendecompositionoftk}
\end{align}
where $\hat{C}_i$ is the projector onto eigenstate $i$.
The first term in the expansion projects onto the ground space as a consequence of the frustration-freeness of the model.
In general, $\tkrandom{U_t}$ is not Hermitian and its left and right eigenspaces may differ.
As a result, each $\hat{C}_i$ may be a non-Hermitian projector.
In all but the simplest circuit geometries, such as the 1d bricklayer circuit (c.f. Appendix~\ref{app:2normdistnonherm}), the non-Hermitian contributions complicate the analysis.
We therefore restrict our attention to circuits with Hermitian layers ($\tkrandom{U_l} =(\tkrandom{U_l})^\dagger$).
For example, one could take any layered circuit and construct a closely related Hermitian circuit by modifying the layers
$\tkrandom{U_l} \rightarrow (\tkrandom{U_l})^\dagger \tkrandom{U_l}$.
We expect our results to hold up to constant rescalings of $t$ even for non-Hermitian layers, although we do not show it here.

To establish a lower bound on $\tau$ we consider the following inequality,
\begin{equation}\label{eq:inequalitychain}
\begin{aligned}
    \sum_{i} e^{-2t\Delta_i} &= \norm{\tkrandom{U_t} - \tkunhaar}^2_{2} \\&\le \norm{\tkrandom{U_t} - \tkunhaar}^2_\diamond \leq \varepsilon^2.
\end{aligned}    
\end{equation}
In the first step we have simply noted that $||\hat{C}_i||^2_2 = 1$ for Hermitian projectors.
The depth $\tau$ is defined as the depth beyond which the diamond norm is less than or equal to $\varepsilon$, which is only possible once the LHS is below $\varepsilon^2$.

To lower bound the LHS of \eqref{eq:inequalitychain}, we consider the variational states $\ket{q} \propto \hat{n}_q \edgeket{\vec{n}}{\sigma}$ created by exciting a Goldstone mode on top of a $+1$ eigenstate of $\tkunhaar$.
At least one such state can be excited in each of $M^k - 2^k$ number sectors -- these are the sectors in which at least one replica carries charge $0<n<M$.
Correspondingly, there must be at least one state with eigenvalue $\lambda \ge e^{-\frac{2 t C}{L^2}}$  in each of these number sectors with $C$ a geometry dependent constant (c.f. Eq.~\eqref{eq:deltageometric}).
We may therefore lower bound the LHS of Eq.~\eqref{eq:inequalitychain} by
\begin{align}\label{eq:grossunderestimate}
    (M^k - 2^k) e^{-\frac{2tC}{L^2}} \le \sum_i e^{-2t\Delta_i}.
\end{align}
Eqs.~\eqref{eq:inequalitychain} and \eqref{eq:grossunderestimate} hold for depths,
\begin{align}
    t \ge \frac{L^2}{2C} \left( \ln(M^k - 2^k) - 2\ln(\varepsilon) \right) .
\end{align}
To better understand this result, we note that the number of charge sectors $M$ is proportional to the volume of the system $M \propto L^d$.
Up to constant factors, the depth at which the random circuit ensemble converges to an $\varepsilon$-approximate $k$-design is therefore lower bounded by
\begin{align}
\label{eq:tau_lowerbound}
    \tau \gtrsim k d L^2 \ln(L) - L^2\ln(\varepsilon) + O(L^{2-kd}).
\end{align}

We now turn to establishing an upper bound on $\tau$ under the assumption that the spectral gap is indeed set by the low energy Goldstone modes: $\Delta \sim C/L^2$.
We follow a similar strategy to that used to establish the lower bound, in particular we calculate the depth $t$ required for the upper bound in Eq.~\eqref{eq:inequalities} to fall below $\varepsilon^2$.
To upper bound the 2-norm we simply assume that all $\globaldim^{2k}$ excited states actually sit at the spectral gap,
\begin{align}
    \sum_i e^{-2t \Delta_i} \le \globaldim^{2k} e^{-\frac{2tC}{L^2}} \le \frac{\varepsilon^2}{\globaldim^{2k}} \label{eq:upperbound2norm}
\end{align}
It is useful to make the dependence on system size explicit by writing $\globaldim = Q^{L^d}$, where $Q = \sum_n \localdim{1}{n}$ is the local Hilbert space dimension. 
Solving for $t$, we find that the channel must converge to an approximate $k$-design for times
\begin{align}
    t \ge \frac{2k L^{d+2}}{C}\ln(Q) - \frac{L^2}{C}\ln(\varepsilon) .
\end{align}
This depth upper bounds the time at which convergence occurs
\begin{align}
\label{eq:tau_upperbound}
    \tau \lesssim k L^{d+2}\ln(Q) - L^2 \ln(\varepsilon).
\end{align}

Compared to the lower bound, the upper bound has gained a factor of $L^d$. 
This can be traced to either of two, extremely loose, upper bounds in the derivation. 
First, in Eq.~\eqref{eq:inequalities} there is a factor of $\globaldim^{2k}$ relating the diamond norm and the 2-norm.
Second, there is a similar factor of $\globaldim^{2k}$ from collapsing all of Hilbert space to the spectral gap.
Either of these alone leads to the extra factor of $L^d$ in Eq.~\eqref{eq:tau_upperbound}, and the combination only changes a constant prefactor.
In the absence of charge conservation, it has recently been shown \cite{harrow2023} that the general diamond norm upper bound used in Eq.~\eqref{eq:inequalities} is, in fact, far too loose when applied to random brick-layer circuit ensembles.
We expect that a similar detailed analysis would reveal that the lower bound in Eq.~\eqref{eq:tau_lowerbound} is tight.

We conclude this section by noting that the definition of approximate $k$-design used in Eq.~\eqref{eq:convergencebound} is a bound on the allowed \emph{additive} error, however some previous work has instead used a stronger definition of approximate design which bounds the \emph{relative} error.
A small relative error implies a small additive error, however a small additive error does not guarantee a small relative error.
A unitary ensemble with additive error $\varepsilon$ may have relative error as large as $\globaldim^{2k} \varepsilon$~\cite{brandao2016a,Schuster_Haferkamp_Huang_2024}.
Demanding a stricter additive error tolerance of $\varepsilon/\globaldim^{2k}$ to meet this definition of approximate design cannot change a lower bound on $\tau$.
Furthermore the conjectural upper bound presented in Eq.~\eqref{eq:tau_upperbound} only changes by a constant prefactor under the modified error tolerance, but retains $O(L^{d+2})$ scaling.

\section{Applications}

Random circuits without number-conservation have found a number of applications owing to their convergence to approximate designs.
In practice, protocols which use a $k$-design on the unitary group $U(\mathcal{H})$ can often instead use an $\varepsilon$-approximate design without significantly changing their outcomes or runtime~\cite{Schuster_Haferkamp_Huang_2024}.
In this section we argue that these results, as applied to circuits without number-conservation, generalize readily to the case of number-conserving circuits.
The argument is straightforward: consider a unitary circuit $U$ drawn from an approximate design with relative error $\varepsilon$.
As noted in Eq.~\eqref{Eq:NDecompofU}, $U$ may be written as a direct sum of $U_n$ acting on individual number-sectors.
Each $U_n$ is therefore drawn from an ensemble which forms an approximate design on $U(\mathcal{H}_n)$ with relative error no greater than $\varepsilon$.
It follows, therefore, that when restricted to the states in $\mathcal{H}_n$ we may directly apply known results for approximate designs on $U(\mathcal{H})$.
For protocols that act on density matrices this argument can be readily extended to classical mixtures across number sectors.
This argument has a number of immediate applications, we highlight in particular:
1) Classical shadows utilizing number-conserving random circuits can estimate $\Tr[\rho\hat{O}]$ with a bias upper bounded by $2\varepsilon\Tr[\hat{O}]$.
2) The purity $\Tr[\rho^2]$ of an unknown state $\rho$ may be estimated from a single copy of $\rho$~\cite{brydges2019} with bias no greater than $2\varepsilon$.
3) Measurement distributions $\mathbb{P}(s|U,\psi) = \abs{\bra{s}U\ket{\psi}}^2$ are close to the Haar distribution.
Since the Haar distribution is presumed difficult to classically sample, measurement distribution from finite depth circuits also be used in demonstrations of quantum hardware.

As an illustrative example, we now focus on the first of these use cases and consider in more detail the application of random circuits to \emph{classical shadows}~\cite{Huang_Kueng_Preskill_2020}.
Classical shadow protocols use the classical measurement results from projective measurements in random bases to efficiently estimate low-rank observables.
Given an unknown state $\rho$ a shadow protocol first performs a random rotation by $U$ then a measurement in the computation basis, resulting in string $b$.
The pair $(U, b)$ is logged into a table of measurement results.
The state $\rho$ is then re-initialized, a new $U$ is sampled, and the procedure repeats.
As an example, after only $O(1)$ such measurements one can construct an unbiased estimate of $\bra{\psi}\rho\ket{\psi}$ for arbitrary pure state $\ket{\psi}$, despite neither $\ket{\psi}$ nor $\rho$ being known at the time of measurement.
More generally, to estimate $\text{Tr}[\rho \hat{O}]$ one needs only take a number of measurements scaling with the rank of $\hat{O}$.

The classical shadow estimator of $\text{Tr}[\rho \hat{O}]$ is constructed assuming that $U$ is drawn from a $3$-design.
This allows us to write an analytical expression for the \emph{reconstruction map}, a linear map which takes the set of measurement outcomes and produces a unbiased estimate of $\rho$, called $\hat{\rho}$.
This is, in turn, used to estimate $\text{Tr}[\rho \hat{O}] \simeq \text{Tr}[\hat{\rho} \hat{O}]$.
One may worry that an estimator using approximate designs may converge more slowly or incur a bias.
By relying on an approximate $3$-design with relative error $\varepsilon$, Ref.~\cite{Schuster_Haferkamp_Huang_2024} demonstrated that the estimator of $\text{Tr}[\rho \hat{O}]$ incurs a bias of at most $2\varepsilon \text{Tr}[\hat{O}]$ and an additional variance bounded by $10 \varepsilon \text{Tr}[\hat{O}]^2$. 

As argued above, these results hold number sector by number sector in our case.
Correspondingly, when constructing a unitary ensemble from $b$-body number-conserving circuits our convergence results then state that this bias bound is applicable only after circuits reach a depth $\tau$ bounded rigorously from below by Eq.~\eqref{eq:tau_lowerbound} and, we conjecture, from above by Eq.~\eqref{eq:tau_upperbound}.

\section{Discussion}\label{sec:Discussion}
We have shown that local random circuits with number conservation are described by a statistical model with an energy gap upper bounded by $\Delta \lesssim 1/L^2$.
We expect the bound on $\Delta$ is tight as it simply reflects the diffusive bottleneck of charge transport in the random circuit; under that assumption we have shown that these circuits form an approximate $k$-design on the full group of number-conserving unitaries in depth $k L^2 \ln(L) \lesssim \tau \lesssim k L^{d+2}$.
This immediately points to a geometric approach to speed up the convergence given a fixed large number of qudits, $V$ -- embed the circuit in higher spatial dimension, where $\tau \lesssim k V^{1+2/d}$.
If long-range gates are available, then a random circuit on an expander graph corresponds to the $d\to\infty$ limit, where $\tau \lesssim k V $ up to logarithmic corrections.

One well-known application of unitary $k$-designs is to measuring the R\'enyi entropy of quantum states \cite{vanenk2012,vermersch2018,elben2018,brydges2019}. 
The entropy can be computed from the statistics of measurements after evolution by Haar random unitaries.
True Haar unitaries are challenging to apply experimentally, but may be approximated by sufficiently deep random local circuits. 
Our results indicate that number conservation, as arises for example in systems of ultracold atoms/molecules, necessitates significantly longer evolution times as compared to what has been discussed in the literature to date. 

Technically, our results rely on the indistinguishability of moments of the Haar measure on the full group of number-conserving unitaries, $U(\mathcal{H}|\hat{N})$, and the group generated by $b$-body circuits, $U_b(\mathcal{H}|\hat{N})$. 
We showed that moments up to $k_c = L^d$ are identical. 
It is clear that $k_c$ is much larger than $L^d$ for $b>2$ or for qudits with more than one state in the maximal or minimal local number sector; either of these conditions leads to super-extensive $k_c$.
Future work could tighten these bounds.

The perspective afforded by spontaneous symmetry breaking in the replica model suggests several immediate generalizations. 
First, random circuits obeying discrete symmetry groups lead to replica models with more intricate but nonetheless gapped discrete symmetry breaking phases. 
These ought to converge as $k$-designs on an $O(L^0)$ time scale, as in the case without symmetry.
Circuits preserving non-Abelian continuous symmetries map to spontaneous symmetry breaking phases with $z=2$ frustration-free Goldstone modes. 
Accordingly, we expect diffusion of non-Abelian charge remains the bottleneck to converge to the group generated by $b$-body circuits. 
In this case, however, this group is a more complicated proper subgroup of $U(\mathcal{H}|G)$ than in the $U(1)$ case~\cite{marvian_restrictions_2022}, and may be distinguishable with low moments. 
For example, random $\text{SU}(d\ge3)$ invariant 2-body circuits fail to generate even $2$-designs of the global $\text{SU}(d\ge3)$ Haar ensemble~\cite{MarvianSUD}.

\begin{acknowledgments}
The authors are grateful to J. Chalker, V. Khemani and A. Potter for stimulating discussions. 
This work was supported by the Air Force Office of Scientific Research through grant No. FA9550-16-1-0334 (M.O.F.) and by the National Science Foundation through the awards DMR-1752759 (S.N.H., M.O.F. and A.C.) and PHY-1752727 (C.R.L.).
\end{acknowledgments}

\appendix

\section{Moment Operator Decomposition From Schur-Weyl Duality}
\label{app:Schurweyl}

In this appendix, we show that the moment operator expansion 
\begin{equation}\label{eq:MomentOperatorExpansion}
\tkuhaar = \sum_{\sigma\tau} V^{\sigma\tau}\ketbra{\sigma}{\tau}
\end{equation}
follows from a well-known expansion of quantum channels in permutation operators. We begin by briefly reviewing that expansion, which is derived in more detail in Ref.~\cite{roberts2017}.

Consider a Hilbert space $\mathcal{H}$, and let $\hat{A}$ be a linear operator on the replicated Hilbert space $\mathcal{H}^{\otimes k}$. Let $\Phi_{k}(\hat{A})$ denote the $k$-fold twirl of $\hat{A}$, defined as
\begin{equation}
    \Phi_{k}(\hat{A}) = \int dU \left(U^{\otimes k}\right)^{\dagger}AU^{\otimes k}
\end{equation}
Due to shift invariance of the Haar measure, $\Phi_{k}(\hat{A})$ commutes with all replicated unitary operators of the form $V^{\otimes k}$. This implies, as a consequence of Schur-Weyl duality, that $\Phi_{k}(\hat{A})$ can be expanded in terms of permutation operators $\hat{\sigma}= \ket{\sigma(\vec{i})}\bra{\vec{i}}$,
\begin{align}\label{eq:SchurWeylConsequence}
    \Phi_{k}(\hat{A}) &= \sum_{\sigma\tau} V^{\sigma\tau} \Tr[\hat{\tau} \hat{A}] \hat{\sigma}
\end{align}
Diagrammatically (taking $k=3$ for simplicity),
\begin{align}
    \Phi_{k}(\hat{A}) &= \int dU \left(
    \begin{tikzpicture}[scale=0.25,baseline={([yshift=-.5ex]current bounding box.center)}]
        \foreach \j in {2,...,0}{
            \draw (-0.25,2*\j+0.5) -- (7,2*\j+0.5);
        }
        \foreach \j in {2,...,0}{
            \draw[fill=blue!40] (1,2*\j) rectangle (2,2*\j+1);
            \draw[fill=yellow!40] (1+4,2*\j) rectangle (2+4,2*\j+1);
        }
        \draw[fill=red!40] (2.8,0) rectangle (4.2,5);
        \node at (3.5,2.5) {$\hat{A}$};
    \end{tikzpicture}
    \right) \\
    &= \sum_{\sigma\tau} V^{\sigma\tau}\left(
    \begin{tikzpicture}[scale=0.25,baseline={([yshift=-.5ex]current bounding box.center)}]
        \foreach \j in {2,...,0}{
            \draw (-0.25,2*\j+0.5) -- (2.25,2*\j+0.5);
        }
        \draw[fill=gray!40] (0.5,0) rectangle (1.5,5);
        \node at (1,2.5) {$\hat{\sigma}$};
    \end{tikzpicture}
    \right) \Tr \left[
    \begin{tikzpicture}[scale=0.25,baseline={([yshift=-.5ex]current bounding box.center)}]
        \foreach \j in {2,...,0}{
            \draw (-0.75,2*\j+0.5) -- (3.75,2*\j+0.5);
        }
        \draw[fill=gray!40] (0,0) rectangle (1,5);
        \draw[fill=red!40] (1.8,0) rectangle (3.2,5);
        \node at (2.5,2.5) {$\hat{A}$};
        \node at (0.5,2.5) {$\hat{\tau}$};
    \end{tikzpicture}
    \right] .
\end{align}

With this result in hand, the claimed expansion for moment operators~\eqref{eq:MomentOperatorExpansion} follows from index rearrangements, which we carry out diagramatically.
Instead of thinking of the channel $\Phi_{k}$ acting on an operator $\hat{A} = \sum_{\vec{i},\bar{\vec{i}}}A_{\vec{i},\bar{\vec{i}}}\ket{\vec{i}}\bra{\bar{\vec{i}}}$, consider the moment operator $\hat{T}_{k}^{U(\mathcal{H})}$ acting on the corresponding state $\ket{A}\equiv\sum_{\vec{i},\bar{\vec{i}}} \hat{A}_{\vec{i},\vec{\bar{i}}}\ket{\vec{i}}\ket{\vec{\bar{i}}}\in\left(\mathcal{H}\otimes\mathcal{H}^{*}\right)^{k}$,
\begin{align}
    \tkuhaar \ket{A} &= \int dU \left(~\begin{tikzpicture}[scale=0.25,baseline={([yshift=-.5ex]current bounding box.center)}]
        \foreach \j in {2,...,0}{
            \draw (-0.25,2*\j+0.5) -- (4,2*\j+0.5);
            \draw (-0.25,2*\j-6+0.5) -- (4,2*\j-6+0.5);
        }
        \foreach \j in {2,...,0}{
            \draw[fill=blue!40] (1,2*\j) rectangle (2,2*\j+1);
            \draw[fill=yellow!40] (1,2*\j-6) rectangle (2,2*\j+1-6);
        }
        \draw[fill=red!40] (3,-6) rectangle (5,5);
        \node at (4,-0.5) {$\ket{A}$};
    \end{tikzpicture}~\right) .
    \label{eq:pictorialtkuhaar}
\end{align}

The state $\tkuhaar\ket{A}$ is related to the operator $\Phi_k(\hat{A})$ by reindexing,
\begin{align}\label{eq:Reindexing}
    \bra{\vec{i},\bar{\vec{i}}} \tkuhaar \ket{A} &= \bra{\bar{\vec{i}}}\Phi_k(\hat{A}) \ket{\vec{i}} .
\end{align}
Diagrammatically, this reindexing manifests as a simple manipulation of the incoming and outgoing indices,
\begin{align}
    \int dU \left(~\begin{tikzpicture}[scale=0.25,baseline={([yshift=-.5ex]current bounding box.center)}]
        \foreach \j in {2,...,0}{
            \draw (-0.25,2*\j+0.5) -- (4,2*\j+0.5);
            \draw (-0.25,2*\j-6+0.5) -- (4,2*\j-6+0.5);
        }
        \foreach \j in {2,...,0}{
            \draw[fill=blue!40] (1,2*\j) rectangle (2,2*\j+1);
            \draw[fill=yellow!40] (1,2*\j-6) rectangle (2,2*\j+1-6);
        }
        \draw[fill=red!40] (3,-6) rectangle (4,5);
        \draw [->,red!40] (4,-5.5) to [out=0,in=270] (6,-3.5);
    \end{tikzpicture}~\right) &\leftrightarrow \int dU \left(~ 
    \begin{tikzpicture}[scale=0.25,baseline={([yshift=-.5ex]current bounding box.center)}]
        \foreach \j in {2,...,0}{
            \draw (0,2*\j+0.5) -- (3,2*\j+0.5);
            \draw (5,2*\j+0.5) -- (9,2*\j+0.5);
        }
        \foreach \j in {2,...,0}{
            \draw[fill=blue!40] (1,2*\j) rectangle (2,2*\j+1);
            \draw[fill=yellow!40] (1+6,2*\j) rectangle (2+6,2*\j+1);
        }
        \draw[fill=red!40] (3,0) -- (3,5) -- (4,5) -- (4,1) -- (5,1) -- (5,5) -- (6,5) -- (6,0) -- cycle;
    \end{tikzpicture}~\right).
\end{align}

Applying the operator expansion of Eq.~\eqref{eq:SchurWeylConsequence} to $\Phi_{k}(\hat{A})$ yields,
\begin{align}
    \Phi(\hat{A}) &= \sum_{\sigma\tau} V^{\sigma\tau}\left(
    \begin{tikzpicture}[scale=0.25,baseline={([yshift=-.5ex]current bounding box.center)}]
        \foreach \j in {2,...,0}{
            \draw (-0.25,2*\j+0.5) -- (2.25,2*\j+0.5);
        }
        \draw[fill=gray!40] (0.5,0) rectangle (1.5,5);
        \node at (1,2.5) {$\hat{\sigma}$};
    \end{tikzpicture}
    \right) \Tr[
    \begin{tikzpicture}[scale=0.25,baseline={([yshift=-.5ex]current bounding box.center)}]
        \foreach \j in {2,...,0}{
            \draw (0,2*\j+0.5) -- (3,2*\j+0.5);
            \draw (5,2*\j+0.5) -- (6.75,2*\j+0.5);
        }
        \draw[fill=gray!40] (1,0) rectangle (2,5);
        \node at (1.5,2.5) {$\hat{\tau}$};
        \draw[fill=red!40] (3,0) -- (3,5) -- (4,5) -- (4,1) -- (5,1) -- (5,5) -- (6,5) -- (6,0) -- cycle;
    \end{tikzpicture}
    ]
\end{align}
Reorganizing the indices in a manner analogous to Eq.~\eqref{eq:Reindexing} leads to the desired result Eq.~\eqref{eq:MomentOperatorExpansion},
\begin{align}
    \tkuhaar \ket{A} &= \sum_{\sigma\tau} V^{\sigma\tau} \left(\begin{tikzpicture}[yscale=0.125,xscale=0.25,baseline={([yshift=-.5ex]current bounding box.center)}]
        \foreach \j in {2,...,0}{
            \draw (-0.75,2*\j+0.5) -- (1,2*\j+0.5);
            \draw (-0.75,2*\j-6+0.5) -- (1,2*\j-6+0.5);
        }
        \draw[fill=gray!40] (0,-6) rectangle (1,5);
        \node at (0.5,-0.5) {$\sigma$};
    \end{tikzpicture}\right) \Tr[
    \begin{tikzpicture}[scale=0.25,baseline={([yshift=-.5ex]current bounding box.center)}]
        \foreach \j in {2,...,0}{
            \draw (0,2*\j+0.5) -- (3,2*\j+0.5);
            \draw (5,2*\j+0.5) -- (6.75,2*\j+0.5);
        }
        \draw[fill=gray!40] (1,0) rectangle (2,5);
        \node at (1.5,2.5) {$\hat{\tau}$};
        \draw[fill=red!40] (3,0) -- (3,5) -- (4,5) -- (4,1) -- (5,1) -- (5,5) -- (6,5) -- (6,0) -- cycle;
    \end{tikzpicture}
    ] \nonumber \\
    &= \sum_{\sigma\tau} V^{\sigma\tau} \left(\begin{tikzpicture}[yscale=0.125,xscale=0.25,baseline={([yshift=-.5ex]current bounding box.center)}]
        \foreach \j in {2,...,0}{
            \draw (-0.75,2*\j+0.5) -- (1,2*\j+0.5);
            \draw (-0.75,2*\j-6+0.5) -- (1,2*\j-6+0.5);
        }
        \draw[fill=gray!40] (0,-6) rectangle (1,5);
        \node at (0.5,-0.5) {$\sigma$};
        \foreach \j in {2,...,0}{
            \draw (2,2*\j+0.5) -- (4,2*\j+0.5);
            \draw (2,2*\j-6+0.5) -- (4,2*\j-6+0.5);
        }
        \draw[fill=gray!40] (2,-6) rectangle (3,5);
        \draw[fill=red!40] (4,-6) rectangle (5,5);
        \node at (2.5,-0.5) {$\tau$};
    \end{tikzpicture}\right) \\
    \tkuhaar &= \sum_{\sigma\tau} V^{\sigma\tau} \left(\begin{tikzpicture}[yscale=0.125,xscale=0.25,baseline={([yshift=-.5ex]current bounding box.center)}]
        \foreach \j in {2,...,0}{
            \draw (-0.75,2*\j+0.5) -- (1,2*\j+0.5);
            \draw (-0.75,2*\j-6+0.5) -- (1,2*\j-6+0.5);
        }
        \draw[fill=gray!40] (0,-6) rectangle (1,5);
        \node at (0.5,-0.5) {$\sigma$};
        \foreach \j in {2,...,0}{
            \draw (2,2*\j+0.5) -- (3.75,2*\j+0.5);
            \draw (2,2*\j-6+0.5) -- (3.75,2*\j-6+0.5);
        }
        \draw[fill=gray!40] (2,-6) rectangle (3,5);
        \node at (2.5,-0.5) {$\tau$};
    \end{tikzpicture}\right) = \sum_{\sigma\tau} V^{\sigma\tau} \ketbra{\sigma}{\tau} .
\end{align}

\section{Generalized Weingarten Functions and the Moment Operator for $U(\mathcal{H})$}
\label{App:Vsigmatau}

In this appendix, we present the explicit formulae for the matrix elements $V^{\sigma\tau}$ of the moment operator 
\begin{align}\label{eq:MomentExpansionApp}
\tkuhaar = \sum_{\sigma\tau} V^{\sigma\tau} \ketbra{\sigma}{\tau}.
\end{align}
in the subspace spanned by the permutation states $\ket{\sigma}$. 
For $k \le \globaldim$, the matrix elements $V^{\sigma\tau}$ are the well-known Weingarten functions.
For $k > \globaldim$, the permutation states are linearly dependent, and the expansion is not unique.
This regime is often ignored in the physics literature.

The matrix elements $V^{\sigma\tau}$ are determined by the overlaps of the permutation states $\ket{\sigma}$ onto which $\tkuhaar$ projects. 
The overlap matrix $D^{\tau\sigma}=\braket{\tau}{\sigma}$ counts cycles in the relative permutation $\sigma \tau^{-1}$, 
\begin{align}
    D^{\tau \sigma} &= \sum_{\vec{i},\bar{\vec{i}}} \braket{\tau}{\vec{i},\bar{\vec{i}}}\braket{\vec{i},\bar{\vec{i}}}{\sigma} \nonumber\\
    &= \sum_{\vec{i},\bar{\vec{i}}} 
    \delta[\bar{\vec{i}} = \tau(\vec{i})]
    \delta[\bar{\vec{i}} = \sigma(\vec{i})] \nonumber\\
    &= \prod_{c \in \textrm{Cycles}(\sigma \tau^{-1})} \globaldim \label{eq:overlapcycles}
\end{align}
Pictorially,
\begin{align}
    \braket{\color{blue}\tau}{\color{red}\sigma} &= 
    \begin{tikzpicture}[scale=0.4,baseline={([yshift=-.5ex]current bounding box.center)}]
        \foreach \j in {5,...,0}{
            \draw (1.75,\j+0.5) -- (3.25,\j+0.5);
        }
        \draw[fill=gray,opacity=0.1] (3,-0.25) rectangle (4.25,6.25);
        \foreach \j in {2,...,0}{
            \draw[red] (3.25,2*\j+0.5) -- (4,2*\j+0.5) -- (4,2*\j+1+0.5) --(3.25,2*\j+1+0.5);
        }
        \draw[fill=gray,opacity=0.1] (0.75,-0.25) rectangle (2,6.25);
        \draw[blue] (1.75,2*2+0.5) -- (1,2*2+0.5) -- (1,2*2+1+0.5) --(1.75,2*2+1+0.5);
        \draw[blue] (1.75,2*0+0.5) -- (1,2*0+0.5) -- (1,2*1+1+0.5) --(1.75,2*1+1+0.5);
        \draw[blue] (1.75,2*0+1+0.5) -- (1.5,2*0+1+0.5) -- (1.5,2*1+0.5) --(1.75,2*1+0.5);
    \end{tikzpicture} = \globaldim^{\vert \text{Cycles}({\color{red}\sigma}{\color{blue}\tau}^{-1})\vert}.\nonumber
\end{align}
Formally, the function $\text{Cycles}(\sigma)$ gives the cycle decomposition of a permutation $\sigma \in S_k$.
Each cycle $c$ is an ordered list of indices $c = (c_1 ... c_{\ell(c)})$ with $c_i \in \{1...k\}$. Graphically, each $c \in \textrm{Cycles}(\sigma)$ corresponds to a closed cycle in the diagram for $\braket{\tau}{\sigma}$.
The diagonal components of the overlap matrix are simply
\begin{align}
    D^{\sigma\sigma} &= \globaldim^k
\end{align}

The coefficients $V^{\sigma\tau}$ are determined by the requirement that $\left(\tkuhaar\right)^2 = \tkuhaar$. 
Plugging in the representation Eq.~\eqref{eq:tkexpansion}, we find this requires
\begin{align}
    V &= VDV.\label{eq:vdv_is_v}
\end{align}
The solution of this system depends on whether $D$ is invertible, that is, whether the $k!$ permutation states are linearly independent. 
For $k \le \globaldim$, all of the $k!$ permutation states are linearly independent and $D$ is invertible~\cite{Collins:2022aa}. 
Hence, Eq.~\eqref{eq:vdv_is_v} simplifies to $V = D^{-1}$.
Readers familiar with Haar averaging will recognize $V^{\sigma\tau} = \text{Wg}(\sigma\tau^{-1}, \globaldim)$ as the Weingarten function~\cite{collins2003,collins2006,Weingarten1978}. 
In this case, the $k!$ permutation states $\ket{\sigma}$ provide a (non-orthogonal) basis for the $+1$ eigenvectors of $\tkuhaar$. 

When $k > \globaldim$ the overlap matrix is not invertible.
Counting the number of $+1$ eigenstates makes the problem obvious; at sufficiently large $k$, $k!$ can exceed the total Hilbert space dimension of the replicated space $\globaldim^{2k}$ so that the $\ket{\sigma}$ states must be linearly dependent. 
However, it is clear how to proceed to construct a projector onto the space \emph{spanned} by the $\ket{\sigma}$ states:
if we can identify a maximal linearly independent subset of the $\ket{\sigma}$, we can drop the others and still obtain an appropriate projector with the matrix of coefficients $V = D^{-1}$ \emph{restricted to the linearly independent set}. 
The appropriate subset is
\begin{align}
    S_k^{\le d} = \{ \sigma \in S_k ~| \textrm{ all}&\text{ increasing subsequences}\nonumber\\&\hspace{15pt}\text{ of }\sigma\textrm{ have length }\le d\} \label{eq:subseqperms}
\end{align}
as identified in Ref.~\onlinecite{Baik:2001aa}, by adapting the representation theory of the unitary group to the moment operators. 
Remarkably, results derived for $k \leq \globaldim$ hold for $k > \globaldim$ if sums over permutation states are restricted to $S_k^{\le \globaldim}$. 
In particular, Eq.~\eqref{eq:tkexpansion} with $V = D^{-1}$ is satisfied when $\sigma, \tau$ are restricted to permutations in $S_k^{\le \globaldim}$.

The trace of the moment operator directly measures the dimension of the $+1$ space. 
We summarize these trace moments for future convenience, 
\begin{align}
\label{eq:haar_tracemom_simple}
    \int dU\,|\Tr U|^{2k} &= \Tr \tkuhaar = k!\quad k \le d 
\end{align}
For the general case where $k > d$ is allowed, the RHS is modified to the number of permutations in $S_k$ for which all increasing subsequences have length $\le d$, 
\begin{align}
    \int dU\,|\Tr U|^{2k} &= |S_k^{\le d}|
\end{align}

\section{Generalized Weingarten Functions and the Moment Operators of $U(\mathcal{H}|\hat{N})$}
\label{app:numberconservinggates}

In this appendix, we generalize the results of App.~\ref{App:Vsigmatau} to the case of number-conserving unitaries.
First, we use combinatoric arguments to compute the trace of $\tkunhaar$, determining the dimension of the $+1$ eigenspace. 
Then, we construct a basis for the space by projecting the parent permutation states $\ket{\sigma}$ onto number sectors and demonstrating that we have exhausted the count obtained by trace arguments.

If $U = \bigoplus_n U_n$ is sampled from the Haar measure $U(\mathcal{H}|\hat{N})$, then each $U_n$ is independently sampled from the Haar ensemble on $U(\mathcal{H}_n)$ (each number block is Haar). 
This allows us to bootstrap results about averaging with respect to the usual Haar ensemble to the number-conserving case.

In the simplest case, the formula Eq.~\eqref{eq:haar_tracemom_simple} is modified:
\begin{align}
\label{Eq:TrUNumberConserving}
    \overline{|\Tr U|^{2k}} &\equiv \int dU\, |\Tr U|^{2k} \\
    &= k!M^k\qquad  k \le d_{n} \, \forall n
\end{align}
where we have introduced an overline notation to indicate averaging with respect to the appropriate ensemble.
Let us show this result:
\begin{widetext}
\begin{equation}
\begin{aligned}
    \overline{|\Tr U|^{2k}} &= \overline{|\Tr \bigoplus_n U_n|^{2k}} \\
     &= \sum_{n_1,\cdots,n_k, \bar{n}_1, \cdots, \bar{n}_k}  
    \overline{ (\Tr U_{n_1})\cdots( \Tr U_{n_k}) (\Tr U^*_{\bar{n}_1})\cdots (\Tr U^*_{\bar{n}_k}) }
\end{aligned}
\end{equation}
\end{widetext}
A term in this sum is only non-zero if the $n$-sectors which appear among the $U$'s match the $n$-sectors which appear among the $U^*$'s, counting with multiplicity. In any configuration of $n_i$, there are
\begin{align}
    \#_n = \sum_i \delta_{n_i, n}
\end{align}
$U$'s in sector $n$. There are then 
\begin{align}
    \binom{k}{\#_0 \cdots \#_{n_m-1}} = \frac{k!}{\prod_n \#_n!}
\end{align}
choices of $\bar{n}_i$ which match the $n$ partition. 
The value of such a term is (assuming $d_n \ge k$):
\begin{align}
\label{eq:prodn_mom_value}
    \overline{|\Tr U_0|^{2\#_0}}\, \overline{|\Tr U_1|^{2\#_1}}\cdots = \prod_n \#_n!
\end{align}
Putting this together,
\begin{equation}
\begin{aligned}
\label{eq:nhaar_mom_value_simple}
     \overline{|\Tr U|^{2k}}  &= \sum_{n_1\cdots n_k} \binom{k}{\#_0 \cdots \#_{M-1}} \prod_n \#_n! \\
     &= \sum_{n_1\cdots n_k} k!\\
     &= k! M^k
\end{aligned}
\end{equation}
as desired.

In the general case, where $d_n < k$ is allowed, the combinatorics are not simple.
The counting of non-zero terms is unchanged, but the value of Eq.~\eqref{eq:prodn_mom_value} is modified:
\begin{align}
    \overline{|\tr U_0|^{2\#_0}}\, \overline{|\tr U_1|^{2\#_1}}\cdots = \prod_n |S_{\#_n}^{\le d}|
\end{align}
where we have used the set notation in Eq.~\eqref{eq:subseqperms}. Putting this together, we find
\begin{align}
\label{eq:nhaar_mom_value_general}
     \overline{|\tr U|^{2k}}  &= \sum_{n_1\cdots n_k} k! \prod_{n} \frac{|S_{\#_n}^{\le d}|}{\#_n!}
\end{align}
As a check, for $k \le d_n \forall n$, we have $|S_{\#_n}^{\le d}| = |S_{\#_n}| = \#_n!$ and the formula reduces to the one in Eq.~\eqref{Eq:TrUNumberConserving}. 

To find a set of states which span the $+1$ eigenspace of $\tkunhaar$, we start with the parent permutation states and project them down by (replicated) number sector:
\begin{align}
    \edgeket{\vec{n}}{\sigma} = \Pi_{\vec{n}} \ket{\sigma}
\end{align}
where the $M^k$ orthogonal projectors $\Pi_{\vec{n}}$ resolve the identity. 
Recalling that $[\Pi_{\vec{n}}, \tkunhaar] = 0$ and that $\tkunhaar$ projects onto a superspace of $\tkuhaar$, we have,
\begin{align}
    \tkunhaar \edgeket{\vec{n}}{\sigma} = \tkunhaar \Pi_{\vec{n}} \ket{\sigma} = \edgeket{\vec{n}}{\sigma}
\end{align}
We note that these $k! M^k$ non-zero states bind the number on the $*$-replicas to that specified by $\vec{n}$ through the permutation $\sigma$ and thus they are jointly diagonal in $\hat{\vec{N}}$ and $\hat{\bar{\vec{N}}}$. 
As expected from their count, these states span $\tkunhaar$; this can also be understood by considering the action of $U$, resolved number block by number block.

The extension of Eq.~\eqref{eq:MomentExpansionApp} to the number-conserving case is
\begin{align}
    \tkunhaar = \sum_{\vec{n},\sigma,\tau}V_{\vec{n}}^{\sigma\tau}\edgeket{\vec{n}}{\sigma}\edgebra{\vec{n}}{\tau}
\end{align}
Shift-invariance requires that $V_{\vec{n}} = V_{\vec{n} }D_{\vec{n}} V_{\vec{n}}$ number sector by number sector, where the overlap matrix within the number sector $\vec{n}$ is 
\begin{align}
    D_{\vec{n}}^{\sigma\tau} = \edgebraket{\vec{n}}{\sigma}{\vec{n}}{\tau} =  \prod_{c \in \textrm{Cycles}(\sigma \tau^{-1})}\delta[n^{c_1} = \cdots = n^{c_k}] d_{n^c}
\end{align}
Explicit combinatorial formulae can be found to invert these matrices  so long as $k \le d_n$ for all $n$. 
In the case where $k > d_n$ for some $n$, the inversion defining $V_{\vec{n}}$ is well-defined by restricting the permutations to $S_k^{\le d_n}$, but does not have a closed form.

\section{Variational Bound on 2-body Brick-Layer Circuits in $d=1$}
\label{app:tl1dbrick}

The arguments of Sec.~\ref{ssec:circuit_gap} are high-level and very general; for completeness, this appendix works through a concrete example for which the circuit gap can be computed exactly. 
We consider a brick-layer circuit built out of 2-body gates acting on a chain with $L$ sites, with $L$ an even integer.
This case can be solved in the sense that $R$ is diagonalized by Fourier modes after mapping onto a classical random walk, as in Eq.~\eqref{eq:randomwalkdef}.
The classical random walk follows from the replacement
\begin{align}
    \tkrandom{U_l} \hat{n}_x \ket{\Psi} &\to \sum_{y} R_{yx} \hat{n}_{y}\ket{\Psi} .
\end{align}
We denote the states on sites of the lattice by rounded kets, $\kket{\cdot}$, to distinguish them from states in the replicated Hilbert space, for which we reserve the standard ket.
In this notation, repeated application of \eqref{eq:randomwalkdef} yields
\begin{align}
    \left(\tkrandom{U_l}\right)^t \hat{n}_x \ket{\Psi} = \sum_{y} \hat{n}_y \ket{\Psi} \bbra{y} R^t \kket{x} 
\end{align}

For a $2$-body brick-layer circuit in a single spatial dimension, $R$ can be written in terms of the states $\kket{x}$ by considering the action of $R$ on each $\hat{n}_x$.
Using periodic boundary conditions, it follows that $R$ is given by
\begin{align}
    R = \frac{1}{4} \sum_{w=0}^{\frac{L}{2}-1} \Big(\kket{2w - 1} + &\kket{2w} + \kket{2w + 1} + \kket{2w +2} \Big)\nonumber\\&\times\Big( \bbra{2w} + \bbra{2w+1} \Big) .
\end{align}
In the following, all sums over $w$ run over the range $\left[0,\frac{L}{2}\right)$, and we omit the limits of the sums.
$R$ is diagonalized by Bloch wavefunctions with a 2-site unit cell,
\begin{align}
    \kket{q_+} &= \frac{1}{\sqrt{L}} \sum_{w} e^{2iqw} \left( e^{-iq}\kket{2w} + e^{iq}\kket{2w+1} \right) \\
    \kket{q_-} &= \frac{1}{\sqrt{L}} \sum_{w} e^{2iqw} \left( \kket{2w} - \kket{2w+1} \right) .
\end{align}
where $q=2\pi m/L$, with $m\in\left[0,\frac{L}{2}\right)$.
These states are eigenstates of $R$:
\begin{widetext}
\begin{align}
    R \kket{q_+} &= \frac{1}{4 \sqrt{L}} \sum_{w} \left(\kket{2w - 1} + \kket{2w} + \kket{2w + 1} + \kket{2w +2} \right)\left( e^{-iq} + e^{iq} \right) e^{2iqw} \\
    &= \frac{\cos(q)}{2\sqrt{L}} \sum_{w} \left( \left(2e^{2iq(w-1)} + 2e^{2iqw}\right)\kket{2w} + \left(2e^{2iqw} + 2e^{2iq(w+1)}\right)\kket{2w+1} \right) \\
    &= \frac{\cos^2(k)}{\sqrt{L}} \sum_{w} e^{2ikw} \left(e^{-ik}\kket{2w} + e^{ik}\kket{2w+1} \right) \\
    &= \cos^2\left(q\right) \kket{q_+} \\
    R \kket{q_-} &= \frac{1}{4 \sqrt{L}} \sum_{w} \left(\kket{2w - 1} + \kket{2w} + \kket{2w + 1} + \kket{2w +2} \right)\left( 1 - 1 \right) e^{2iqw} \\
    &= 0 
\end{align}
\end{widetext}

In the case where $L$ is not divisible by four, these eigenstates are linearly independent and diagonalize $R$. However, if $L$ is divisible by four, then $q=\pi/2$ is an allowed momentum, and it is straightforward to check that $\kket{\frac{\pi}{2}_{+}}$ is equal to $\kket{\frac{\pi}{2}_{-}}$, up to a global phase. Therefore, an eigenvector of $R$ at momentum $\pi/2$ is missing. This can be fixed by noting that the following two states are eigenstates of $R$ with eigenvalue $0$ and momentum $\pi/2$:
\begin{align}
    \kket{\phi_{1}}&=\sum_{w}e^{i\pi w}\left(\kket{2w} + \kket{2w+1}\right)\\
    \kket{\phi_{2}}&=\sum_{w}e^{i\pi w}\left(\kket{2w}-\kket{2w+1}\right)
\end{align}
These states, in combination with the states $\kket{q_{+}}$ and $\kket{q_{-}}$ $(q\neq \pi/2)$, diagonalize $R$.

In either case, $R$ has a set of degenerate eigenvectors with eigenvalue zero, and another set $\kket{q_{+}}$ with eigenvalues $\cos^{2}(q)$. 
The action of $R$ on its eigenstates can be used to study $\tkrandom{U_l}$.
Consider the state
\begin{align}
\ket{q_+} &= \frac{1}{\sqrt{L}} \sum_w e^{2iqw} \left(e^{-iq}\hat{n}_{2w} + e^{iq}\hat{n}_{2w+1}\right)\ket{\Psi} .
\end{align}
We can then, by the relation in \eqref{eq:randomwalkdef}, equate the eigenvalues of $R$ to the bound~\eqref{eq:circuitboundfractioncomplex},
\begin{align}
    \frac{\vert R \kket{q_+} \vert^2}{\bbrakket{q_+}{q_+}} = \frac{\vert \tkrandom{U_l} \ket{q_+}  \vert^2}{\bra{q_+}\ket{q_+}} = \cos^4(q) \le e^{-2\Delta}
\end{align}
For small $q$ we expand $\cos^4(q) \simeq e^{-4q^2}$, and find the bound $\Delta \le 2 q^2$, consistent with the results of Sec.~\ref{ssec:circuit_gap}.

\section{2-norm distance of non-hermitian operators and the d=1 bricklayer circuit}
\label{app:2normdistnonherm}

For general circuit geometries, the moment operator for a depth $t$ circuit, $\tkrandom{U_t}$, is represented by a non-hermitian matrix with distinct left- and right- eigenvectors.
The 2-norm distance between $\tkrandom{U_t}$ and the Haar moment operator $\tkunhaar$ acquires contributions from the hermitian and anti-hermitian components of $\tkrandom{U_t}$.
In general, it is difficult to bound the geometry dependent anti-hermitian contribution, thought we show here that it is negligable in the case of the d=1 bricklayer circuit.

To begin, we note that shift invariance of the Haar ensemble allows us to simplify the 2-norm distance into the difference of 2-norms,
\begin{align}
    \norm{\tkunhaar - \tkrandom{U_t}}^2_2 &= \norm{\tkrandom{U_t}}^2_2 - \norm{\tkunhaar}^2_2 .
\end{align}
Furthermore, by changing basis to make $\tkrandom{U_t}$ upper triangular it is straightforward to show that $\norm{\tkrandom{U_t}}^2_2$ may be broken into a sum over squared eigenvalues and an off-diagonal contribution,
\begin{align}
    \norm{\tkrandom{U_t}}^2_2 &= \sum_{i} |\lambda_i|^{2t} + \frac{1}{2} \norm{\tkrandom{U_t} - \left(\tkrandom{U_t}\right)^\dagger}^2_2 \label{suppeq:twonormantiherm}
\end{align}
with $\lambda_i$ eigenvalues of the single layer moment operator.

We now turn to the special case of the d=1 bricklayer circuit.
The moment operator for this circuit geometry may be written as an alternating product of two hermitian projectors corresponding to the even and odd sublayers, $\tkrandom{U_t} = \left(P_{\text{odd}} P_{\text{even}}\right)^t$.
As a result, $\tkrandom{U_t}$ has real eigenvalues and the anti-hermitian contribution to \eqref{suppeq:twonormantiherm} greatly simplifies,
\begin{align}
    \norm{\tkrandom{U_t} - \left(\tkrandom{U_t}\right)^\dagger}^2_2 &= \Tr[\tkrandom{U_{2t-1}}] - \Tr[\tkrandom{U_{2t}}]\\
    &= \sum_i \lambda_i^{2t-1} \left( 1 - \lambda_i \right)
\end{align}
Plugging this and \eqref{suppeq:twonormantiherm}  back into the expression for the 2-norm distance gives a final expression for the depth $t$ d=1 bricklayer circuit in terms of its single-layer eigenvalues,
\begin{align}
    \norm{\tkunhaar - \tkrandom{U_t}}^2_2 &= \frac{1}{2} \sum_{\lambda_i < 1} \lambda_i^{2t-1}(1 + \lambda_i) .
\end{align}

\bibliography{mybib}

\end{document}